\documentclass[useAMS,usenatbib]{almamemo}
\pdfoutput=1
\usepackage{amsmath}
\usepackage{microtype}
\usepackage{mathptmx}
\usepackage{graphicx}
\usepackage{booktabs}
\usepackage{tabularx}
\usepackage[colorlinks=true,citecolor=blue]{hyperref}
\usepackage[]{subfig}
\newcommand{\unit}[1]{\mathrm{#1}}
\newcommand{\unitp}[2]{\ensuremath{\mathrm{#1}^{#2}}}

\newcommand{\bzrdate}{2009-07-24 16:33:15 +0100}
\newcommand{\bzrrevno}{207}

\volume{588}
\title[]{ALMA Memo \#588\\
Inference of Coefficients for Use in Phase Correction II:\\ Using the
Observed Correlation Between Phase and Sky Brightness Fluctuations
}

\author[]{B. Nikolic\\Mullard Radio
  Astronomy Observatory, Cavendish Laboratory, Cambridge CB3 0HE, UK
  \\\url{email:b.nikolic@mrao.cam.ac.uk}
  \\\url{http://www.mrao.cam.ac.uk/~bn204/}}

\date{24th July 2009}
\pagerange{\pageref{firstpage}--\pageref{lastpage}; } \pubyear{2009}

\begin{document}
\label{firstpage}
\maketitle

\begin{abstract}
By observing bright and compact astronomical sources while also taking
data with the 183\,GHz Water Vapour Radiometers, ALMA will be able to
measure the `empirical' relationship between fluctuations in the phase
of the astronomical signal and the fluctuations of sky brightness
around 183\,GHz. Simulations of such measurements assuming only
thermal noise in the astronomical and WVR receivers are presented and
it is shown that accurate determination of the empirical relationship
should be possible in a relatively short time. It is then proposed
that the best way of using these empirical coefficients is to include
them as a constraint on a physical model of the atmosphere -- this
allows them to be used for longer period of time, increasing the
efficiency of observing.  This approach fits naturally into the
analysis framework presented in the previous memo, which has now been
extended to implement it. The technique is illustrated via simulations
and on a short data set collected at the SMA.
\end{abstract}

\section{Introduction}

The Atacama Large (sub-)Millimetre Array (ALMA) will calibrate and
correct the effects of the Earth's atmosphere on the phase of the
observed radiation by a combination of two techniques:
\begin{enumerate}
  \item Frequent (every $\sim20$--200\,s, depending on the conditions
    and the scientific programme) fast-switching observation of quasars
  \item Water Vapour Radiometry (WVR) using dedicated 183\,GHz
    radiometers to be installed on all 12\,m-diameter antennas
\end{enumerate}
The first technique requires suspension of science observing for a few
seconds but allows a direct observation of the excess path due to the
atmosphere between any two antennas. The second technique will be
applied continuously, probably even during small changes in observing
direction, but requires a model which will translate WVR outputs to
estimates of excess path.

A model for calculating the coefficients that translate WVR outputs to
estimates of path fluctuation can be constructed using the physical
properties of the atmosphere and its constituents. For example, the
previous memo in this series describes an extremely simple model
\citep{ALMANikolic587} of just a single, thin, layer of water
vapour. Whatever the model is that is chosen, there will however be
some errors which will limit the accuracy with which the phase
correction coefficients are estimated. These errors can be divided in
two categories:
\begin{itemize}
  \item Uncertainties in the parameters of the model, e.g., height and
    temperature of the water vapour in the atmosphere
  \item Incorrect physical assumption or omission of important
    effects, e.g., in a thin water-vapour layer model, it may be the
    assumption of \emph{thinness\/} that introduces a significant
    error
\end{itemize}

When the commissioning of ALMA begins, it will be possible to asses
the accuracy of such models by observing quasars and examining how
well the predicted path fluctuations correspond to those measured from
the visibilities of the quasar. Clearly, such testing this will allow
for better models with most of the relevant physics to be selected. 

In this memo, I consider a way of taking this a step further and using
such test observations to constrain the \emph{model parameters\/} as
well model section. There are two reasons why this may be practical
and useful with ALMA:
\begin{enumerate}
  \item ALMA will have high sensitivity and has been designed to
    efficiently do the fast-switching observations. This means that
    observations of quasars which allow both the interferometric path
    and WVR sky brightness to be recorded can be done with only a
    small interruption to science observing. Therefore the relevant
    `test' data can be obtained cheaply and continuously and should be
    used to best advantage
  \item Such observations of quasars allow empirical measurement of
    the relationship between outputs of the WVRs and interferometric
    phase. Clearly this information must be very relevant for any
    model which tries to predict this same relationship
\end{enumerate}

The main part of the memo begins with an analysis
(Section~\ref{sec:obssacc}) of how well the relationship between
fluctuations of WVR outputs and interferometric path can be estimated
in practice, with both a single-base line system and a full 50 antenna
ALMA array. Then, I will show (Section~\ref{sec:results}) that such
empirical estimates can be used to better constrain the parameters of
an atmospheric model, taking in this case the simple model previously
proposed. Finally the procedure will be illustrated
(Section~\ref{sec:smaillustration}) on some test data collected at the
Submillimeter Array with prototype ALMA WVRs.

\section{Linearisation of fluctuations}
\label{sec:linearisation}

For the application of phase correction to an array such as ALMA, we
are primarily interested in fluctuations (over timescales from one
second to about 200\,seconds) of path to each of the antenna rather
then the absolute value of the path. The reasons for this are:
\begin{itemize}
  \item An interferometer such as ALMA is not sensitive to absolute
    changes in path, only the relative paths between its antennas
  \item The fast-switching phase calibration can easily remove any
    phase errors on time scales longer than about 200\,s
  \item Instrumental drifts in both the interferometer and the WVRs
    may become dominant sources of phase error in ALMA timescales
    longer than about 200\,s
\end{itemize}

The fluctuations in path $\delta L$ will be estimated from the
fluctuations of the WVR outputs, $\delta T_{{\rm B},i}$ according to:
\begin{equation}
      \delta L \approx \sum_i w_i \frac{{\rm d} L}{{\rm d} T_{{\rm B},i}} \delta
      T_{{\rm B},i}
\end{equation}
where $w_i$ is the weighting of the channels and ${\rm d}L/{\rm d}
T_{{\rm B},i}$ are the `phase-correction coefficients'. The
coefficients are predicted by atmospheric models and, when observing a
strong source of known structure, can be estimated observationally (as
described in the next section). 

Inherent in this expression is the assumption that $\delta L$ and
$\delta T_{{\rm B},i}$ are linearly related. This certainly will not
be the case for large changes in conditions as the 183\,GHz is close
to saturation even in the dry conditions prevalent at the ALMA
site. However for small fluctuations and short time scales, this
should be a reasonable approximation.

Throughout this memo, it is assumed that the timescales and baselines
are short enough that this linear relationship holds. This means one
can consider ${\rm d}L/{\rm d} T_{{\rm B},i}$ as constant during
intervals of interest. If the elevation of the antennas are reasonably
constant, it is expected that this will be true on timescale a few
minutes or so, which is sufficiently long for the technique described
here. If the astronomical source is setting or rising quickly however,
this may not be true, and care should be taken to take into account
second-order effects.

\section{How accurately can we estimate phase correction coefficients
  observationally?}

\label{sec:obssacc}

\begin{figure}
  \includegraphics[clip,width=0.99\linewidth]{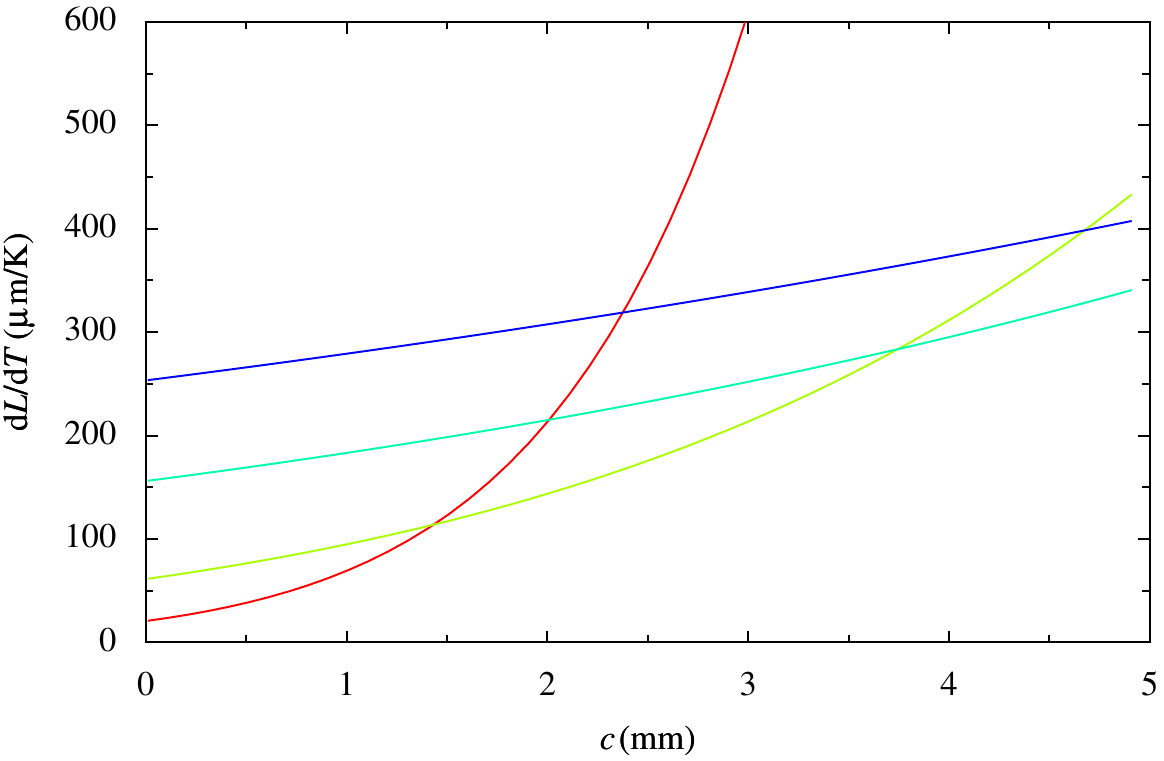}
  \caption{Model values of the phase correction coefficients, ${\rm
      d}L/{\rm d} T_{{\rm B},i}$, for a wide range of line-of-sight
    water vapour columns ($c$), calculated for the ALMA production
    radiometer. Colours represent the four channels of the WVRs: red
    is channel 1 (the inner-most channel), yellow-green is channel 2,
    turquoise is channel 3 and blue channel 4 (the outer-most
    channel).}
  \label{fig:coeffsmodel}
\end{figure}

There are two unavoidable (but also, easy to estimate) sources of
error in estimating the phase correction coefficients directly from
the correlation between the observed fluctuations in path and sky
brightness. These are the thermal-like errors arising in the mixers
and front-end amplifiers of both the astronomical and of the WVR
receiving systems.

The noise in the astronomical system will produce an error in the
observed visibility between each pair of antennas, and therefore in
the inferred path fluctuation between them. The magnitude of this
error will depend on:
\begin{enumerate}
  \item Receiver noise temperature
  \item Strength of the calibration source
  \item Number and size of antennas in the array
\end{enumerate}
In the regime when path errors are small, they are closely related to
the estimated \emph{flux\/} sensitivity of the calibration observation
as follows.  The error in the estimated \emph{phase\/} from each
visibility is approximately $\delta \phi \sim \delta S/S$ where
$\delta S$ is the noise in measured flux, $S$ is the calibration
source strength and $\delta \phi$ is in radians.

The sensitivity of radio interferometers is given, e.g., by
\cite{WilsonTORA}:
\begin{equation}
  S_\nu=\frac{2 M k  T_{\rm sys}'}{A_{\rm e} \sqrt{2 N t \Delta \nu}}
\end{equation}
Where the symbols have following meaning:
\begin{description}
  \item[$N$] Number of correlations, $n(n-1)/2$ for a full array of
    $n$ antenna
  \item[$M$] Digitisation, correlation, efficiency factor, $\sim 1$
  \item[$T_{\rm sys}'$] System temperature, including atmospheric
    absorption
  \item[$A_{\rm e}$] Effective antenna area, $90\,\unitp{m}{2}$ for the
    $12\,\unit{m}$ ALMA antennas
  \item[$t$] Integration time
  \item[$\Delta \nu$] Bandwidth
\end{description}
and final sensitivity is for a single polarisation. Since we are
interested in measurement of \emph{antenna\/} phases, the relevant
number of correlations is $N=n-1$, i.e., 49 for the case of the main
ALMA array. If we take roughly representative parameters for phase
calibration at 90\,GHz, i.e., $M=1$, $N=49$,$T_{\rm
  sys}'=50\,\unit{K}$, $A_{\rm e}=90\,\unitp{m}{2}$, $t=1\,\unit{s}$,
$\Delta \nu=16\,\unit{GHz}$ (to account for the fact that both
polarisations will be measured) we find $\delta
S_\nu=1.2\,\unit{mJy}$. Therefore to achieve 1 degree phase solution
accuracy, approximately a $70\,\unit{mJy}$ calibration source needs to
be observed. In the remainder of this paper, we will assume that the
strength of the calibration sources are indeed observed and that the
resulting path accuracy is $10\,\micron$. Such phase error would lead
to only a small contribution to de-correlation even at the shortest
operational wavelengths for ALMA.

For initial tests of the phase correction techniques at ALMA, it is
likely only a single baseline will be available. In this case the
error on the phase measurement will be approximately 35 times greater
and so sources of around 2.5\,Jy should be observed for the results
here to be applicable.

The intrinsic, thermal-like, errors on the measurements by the WVRs
will be in the range of $0.07\,\unit{K}$ to $0.1\,\unit{K}$
root-mean-square in one second of integration time, depending on the
WVR channel. Although the goal of this memo is to find the constraints
on the phase correction coefficients, simple approaches
\citep[e.g.,][]{ALMANikolic587} can easily yield estimates which are
good enough to translate the above \emph{errors\/} in terms of Kelvin
into corresponding path errors, at the frequency at which the
calibration source is observed. 

Using the simple single layer model presented previously
\citep{ALMANikolic587}, phase correction coefficients as a function of
water-vapour column only are shown in
Figure~\ref{fig:coeffsmodel}. This figure therefore shows
approximately how the thermal errors in the WVRs will translate to
path errors for a range of conditions at the ALMA site. In the median
conditions at the site, which correspond to water vapour column of
1\,mm, the expected error is in the range of less than $10\,\micron$
to about $30\,\micron$, depending on the WVR channel. Therefore, in
the results presented above, we will do the analysis with assumed
errors of $10\,\micron$, $20\,\micron$ and $50\,\micron$.

There are likely to be other sources in error which have been
neglected here, for example:
\begin{enumerate}
  \item Instrumental drift of interferometer path
  \item Drift in the WVRs sky temperature (specified to be less than
    $0.1\,\unit{K}$ in 10 minutes)
\end{enumerate}
Although we neglect these in the present work, they may turn out to be
a significant limitation in practice. We note however that we limit
ourselves to observations around of length of ten minutes or shorter
which should restrict the instrumental drifts. 

\subsection{Simulating fluctuations}

\begin{figure*}
  \includegraphics[clip,width=0.99\linewidth]{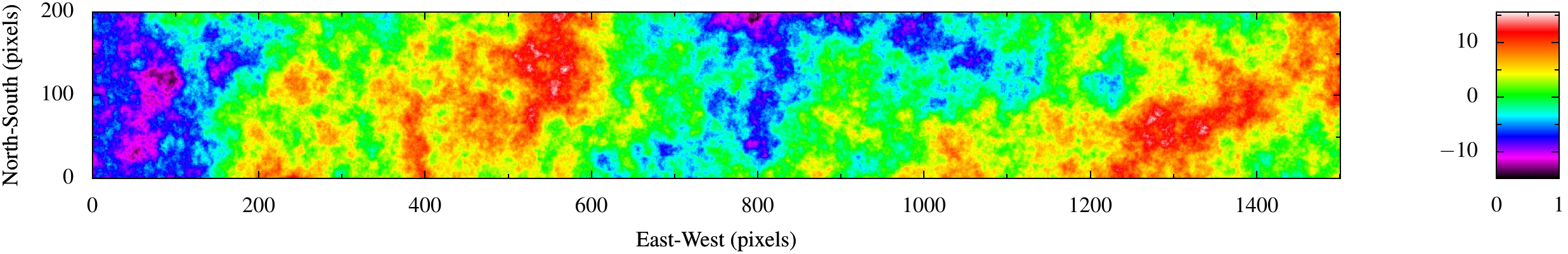}
  \caption{A one-eight subsection of the turbulent phase screen used
    in the simulation of empirically determining the correlation
    between temperature brightness and phase fluctuations. The scale
    is in arbitrary units, as the screen is later re-scaled so that
    the fluctuation on a 300\,m baseline is 200\,\micron. }
  \label{fig:screen}
\end{figure*}

Beside the intrinsic errors on measurements of the interferometric
path fluctuation and the sky brightness temperature, the accuracy with
which the correction coefficients can be estimated will also depend on
the magnitude of the fluctuations and the length and number of data in
the observation. 

In order to quantify the typical magnitude of fluctuations, we have
made simulations using the methods presented in detail in earlier
memos \citep{ALMANikolic573,ALMANikolic582}. The phase screen
generated for this study is shown in Figure~\ref{fig:screen} and
corresponds to a $100\,\unit{m}$ thick turbulent layer in the
atmosphere. Fluctuations are scaled so that they are $200\,\micron$ on
a $300\,\unit{m}$ baseline, i.e., they correspond to the median
conditions at the ALMA site.

\begin{figure}
  \includegraphics[clip,width=0.99\columnwidth]{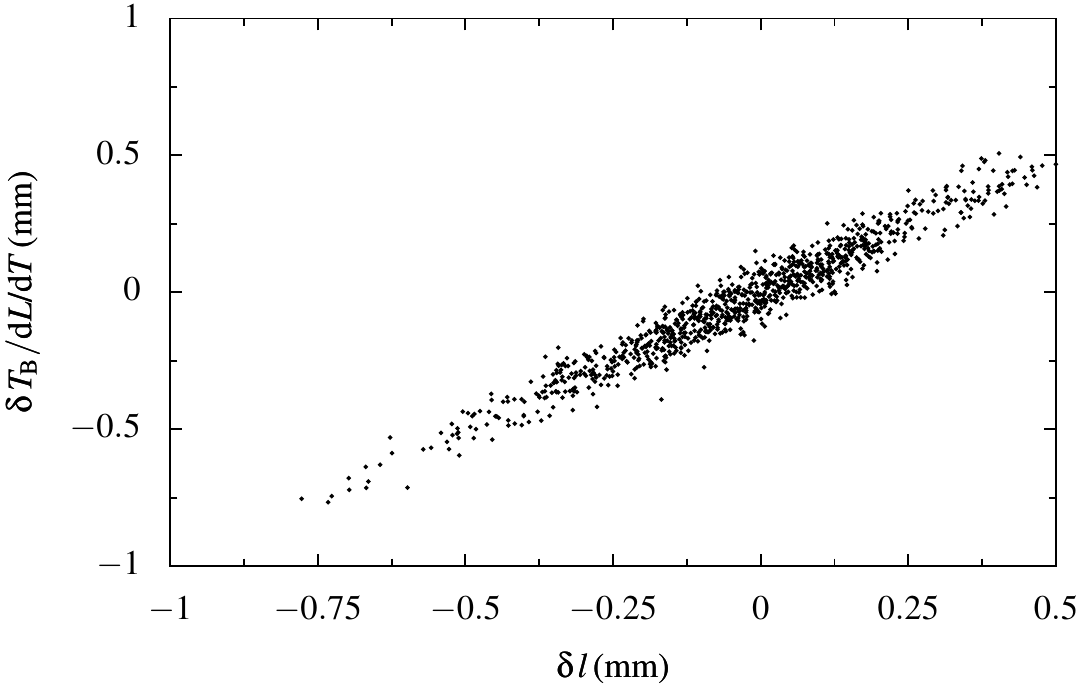}
  \caption{Example of simulated data, with errors in both
    coordinates. Data are on an approximately 300\,m baseline and
    sampled every one second for 1000 seconds. Error on path is
    assumed to be 10\,\micron\ and error on sky brightness equivalent
    to 50\,\micron. The phase screen used is shown in
    Figure~\ref{fig:screen}.  }
  \label{fig:scatterfitex}
\end{figure}

A typical set of simulated data is shown in
Figure~\ref{fig:scatterfitex}, illustrating the distribution of points
due to the statistics of atmospheric fluctuations, and the assumed
noise on path and sky brightness temperatures.

We will consider simulations of two types of observations:
\begin{enumerate}
\item Calibration observations of that are interleaved with science
  observations. We assume calibration observation are 1 second long
  and the intervals between them are in the range of 2 to 200 seconds,
  with total length of sequence of 1000 seconds
\item Dedicated calibration observations of length up to 200 seconds
\end{enumerate}

Finally, in order to estimate the accuracy with which observations can
be used to constrain the phase correction coefficients, we repeat the
simulations 100 times to create a Monte-Carlo simulation. 

\subsection{Finding the best-fitting slope}
\label{sec:linefitting}

Since we are assuming the approximate linear relationship between
fluctuation in path and the sky brightness, we are only interested in
finding the slope of the best fitting line which connects the two
quantities. By design of the ALMA WVR system, the errors on the
observed path and the observed sky brightness are approximately the
same magnitude. As a result, the computation of the best-fitting line
is slightly more involved than the usual case in which the only
significant errors are in one (`$y$') coordinate.

We adopt the following notation: $\hat{y}$ and $\hat{x}$ represent the
true values of path and temperature and they are assumed to be related
exactly by:
\begin{align}
  \hat{y}=a\hat{x}+b
\end{align}
where $a$ is the slope coefficient that we seek to find. The observed
values $y$, $x$ differ from the true values by random errors that are
normally distribution with known variance, i.e.,
$y=\hat{y}+\epsilon_y$ where $\epsilon_y\sim N(0,\sigma_y)$ and
similarly for $x$, $x=\hat{x}+\epsilon_x$ where $\epsilon_x\sim
N(0,\sigma_x)$. The values of $\epsilon_x$ and $\epsilon_y$ can in
practice estimated \emph{a-priori\/} using the calculations from the
beginning of this section.

A relatively simple, maximum-likelihood, procedure for finding the
best fitting line is given by \cite{PressNRC}.  This involves
calculating the approximate negative log-likelihood of the observation
given model parameters:
\begin{align}
  F(a,b) = \frac{\sum_i \left(y_i - a x_i -b\right)^2}{\sigma_y^2+a^2 \sigma_x^2}
\end{align}
and then simply minimising function $F(a,b)$ with respect to the two
parameters. We implement this using the Levenberg-Marquardt algorithm
which produces an estimate of $a$ and of the formal error $\sigma_a$
although this will only be approximately correct. The implementation
is available in {\tt BNMin1}, version 1.6 \citep{BNMin1}.  (A complete
solution to the fitting is given by \cite{GullLine1989}, but has not
been used here).

The formal error derived from the fitting procedure, $\sigma_a$, is
used for combining estimates from different baselines in the array
(see below), but the final quoted accuracy is always determined by a
Monte-Carlo simulations from 100 realisation of the simulated data
sets.

\subsection{Single baseline}
\label{sec:single-baseline-res}

\begin{figure}
  \includegraphics[clip,width=0.99\linewidth]{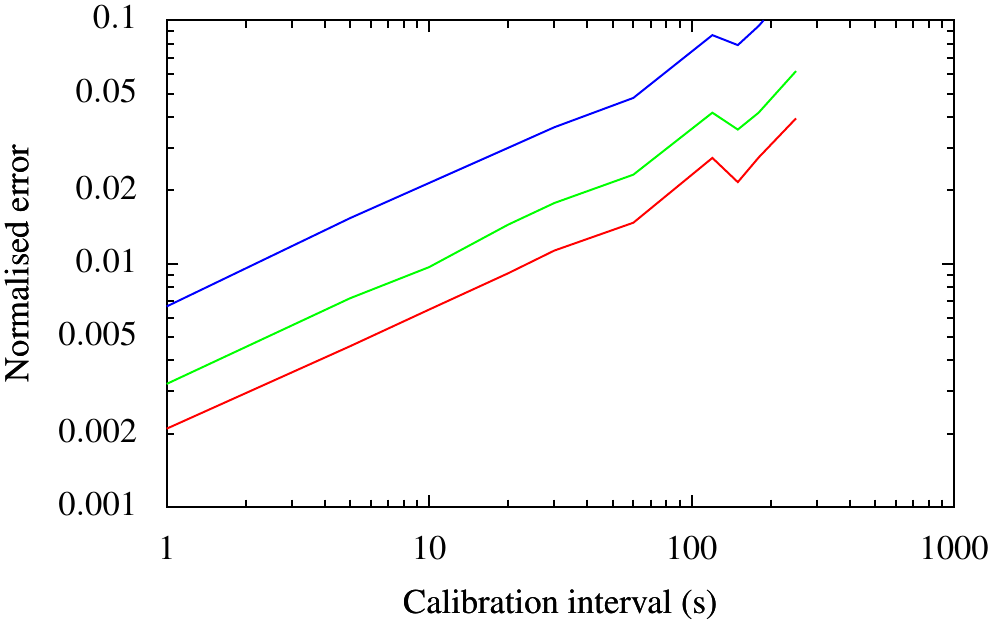}
  \caption{Single baseline measurement: Normalised error on the estimated slope of the correlation
    between phase and sky-brightness fluctuations as a function of the
    time interval between calibration observations. It is assumed the
    interferometric path measurement has a normally-distributed error
    of 10\,\micron\ and the radiometric sky brightness measurement has
    a normally distributed error of 10\,\micron\ (red line),
    20\,\micron\ (green) and 50\,\micron\ (blue).}
  \label{fig:singlebase-interv}
\end{figure}

\begin{figure}
  \includegraphics[clip,width=0.99\linewidth]{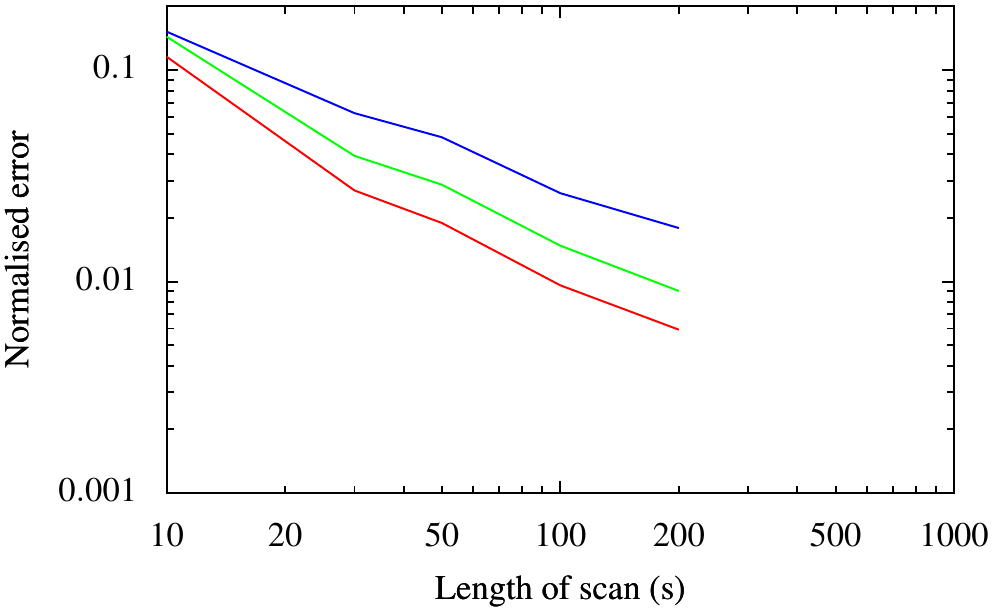}
  \caption{Single baseline measurement: Normalised error on the
    estimated slope of the correlation between phase and
    sky-brightness fluctuations as a function of the length of the
    specialised scan for estimating it. It is assumed the
    interferometric path measurement has a normally-distributed error
    of 10\,\micron\ and the radiometric sky brightness measurement has
    a normally distributed error of 10\,\micron\ (red line),
    20\,\micron\ (green) and 50\,\micron\ (blue).}
  \label{fig:singlebase-scan}
\end{figure}

As mentioned previously, in the initial stages of ALMA commissioning
and testing of the WVR-based phase correction, it is likely that we
will be working with a single baseline configuration.  For this the
reason, we have performed an analysis of how well the phase correction
coefficients can be estimated observationally with just a single
baseline between two 12\,m diameter antennas.

The results are shown in Figures \ref{fig:singlebase-interv} and
\ref{fig:singlebase-scan}. The first figure
(\ref{fig:singlebase-scan}) is for the observing strategy which
consists of interleaved observations of calibrator and science
targets. The vertical axis represents the normalised error in the
estimated phase correction coefficient and the horizontal axis the
interval between the phase calibration scans. The total observation
time in all cases is 1000\,seconds. As described above, the errors due
thermal noise in the WVRs will depend on the conditions of observation
and channel of the WVRs that is analysed, so we only show three
representative cases by the three coloured lines on the plot.

As can be seen in the figure, for the single baseline configuration we
can expect errors in the range of 1\% to 5\% if the interval between
calibration observations is in the range 20--100\,s. One possibility
therefore is to carry out experiments to \emph{simultaneously} track
the phase of a few quasars at a range of elevations. This is more
efficient than doing these observations sequentially since allowing
for time to elapse between taking data points results in greater phase
changes and therefore higher accuracy.

The results shown in Figure~\ref{fig:singlebase-scan} are for a
dedicated calibration scan, during which the phase calibration source
is continuously observed. The vertical axis again represents the
normalised error on the measured slope, while the horizontal axis
represents the length of the calibration scan. The three lines are
again calculated for representative values of errors due to thermal
noise in the WVRs.

It can be seen that for a range of parameters, the errors in inferred
phase correction in this procedure are larger than in the procedure
using interleaved observations of calibrators. The reason is again
that interleaved observations allow greater changes in path to
occur. Never the less, for scans longer than about 60 seconds, errors
of the order of 5\% or less can be expected.

\subsection{Full array}

We next consider a more typical situation for ALMA, in which the whole
bi-lateral array of 50 12\,m-diameter antennas is used. In this
situation, an estimate of the phase correction coefficient can be
derived from each baseline, but with the caveat that determinations
from individual baselines will depend on baseline length and will not
be statistically independent.  This can be tackled easily within the
simulation already developed for the single baseline as follows.

First, it should be noted that when calculating the antenna-based
complex gain solution from quasar calibration observations there is a
degree a freedom that is always unconstrained by the observation,
i.e., the "common-mode". This missing information is usually
regularised by either assuming that the phase errors on one of the
antennas are always zero, or that the mean phase error of all antennas
is zero.

The measurements from the WVRs are however intrinsically absolute
measurements in which this common-mode is still present. Therefore
correlating complex gain solution of an antenna with the WVR signal
could give quite misleading results, if this common mode signal is
significant. This is likely to be the case even in the extended
configurations of ALMA, because of the tendency of atmospheric
turbulence phenomena to increase in magnitude on longer length-scales.

One can get around this problem by considering the difference in
signals between pairs of radiometers (thereby removing the
common-mode) and correlating it with the difference in phase between
the gains of the two antennas. This procedure is most naturally done
by skipping the step of calculating the antenna phase gain solutions
completely, since individual visibilities are in the present case
nothing more than direct measurements of difference in phase of gains
of pairs of antennas. Therefore in the calculation below we work
directly with visibilities recorded on each baseline and correlate
with the differences in the radiometer signals.

The simulation method was as follows.  Antenna path fluctuation were
simulated for each of the antennas for the duration of the
observation.  The estimated noise is then added to the simulated
antenna paths, and to the simulated radiometer outputs. Each of the
baseline is then formed and slope of correlation estimated
individually, together with the formal error on the slope,
$\sigma_a$. Finally estimates on each baseline are combined by:
\begin{equation}
  \bar{a}= \frac{1}{\sum_j 1/\sigma^2_{a,j}} \sum_j a_j/\sigma^2_{a,j}  
\end{equation}
where the sum over $j$ runs over all baselines. It is essential to
weight the estimates by their formal errors, because the errors on
short baselines are much larger than on the long baselines.

Note here we made an inherent assumption that the coefficients on all
antennas are the same. There are two reasons we can foresee now that
this will may not be the case:
\begin{enumerate}
  \item Different atmospheric conditions at different antennas,
    especially in the most extended configurations when antennas will
    be separated by up to 15\,km and have significantly different
    altitudes 
  \item Different filter responses of WVR units
\end{enumerate}
If these differences are found to be significant in practice they can
be included in the modelling although at a cost of a significant
increase in complexity.

\begin{figure}
  \includegraphics[clip,width=0.99\linewidth]{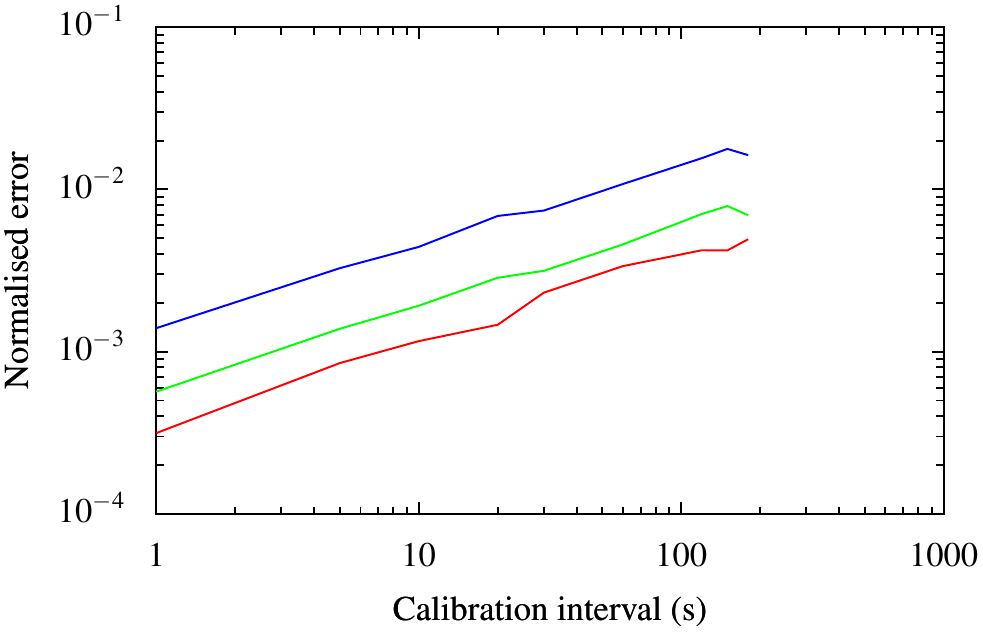}
  \caption{Full 50 antenna ALMA array measurement: Normalised error on
    the estimated slope of the correlation between phase and
    sky-brightness fluctuations as a function of the interval between
    consecutive calibration scans. The total scan is assumed to be
    1000\,s long.It is assumed the interferometric path measurement
    has a normally-distributed error of 10\,\micron\ and the
    radiometric sky brightness measurement has a normally distributed
    error of 10\,\micron\ (red line), 20\,\micron\ (green) and
    50\,\micron\ (blue). The whole array of 50\, ALMA antennas is
    simulated.}
  \label{fig:array-interv}
\end{figure}

\begin{figure}
  \includegraphics[clip,width=0.99\linewidth]{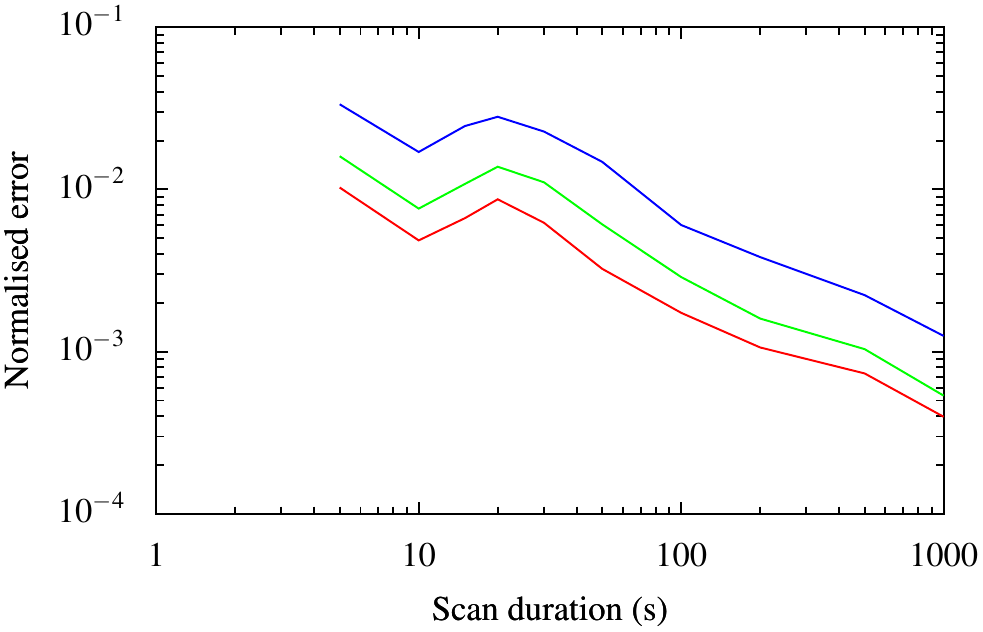}
  \caption{As Figure~\ref{fig:array-interv}, but for  single
    continuous calibration observations of varying length.  This
    figure is also for the full 50 antenna ALMA array.}
  \label{fig:array-scan}
\end{figure}

The results for the full array simulations are shown in Figures
\ref{fig:array-interv} and \ref{fig:array-scan}, for the interleaved
and continuous quasar observations respectively. With the full array,
the accuracy of the measurements is of course much higher than on a
single baselines and accuracies of better than 1\% should be easily
achievable unless significant errors in addition to those introduced
by thermal noise become apparent.

\section{Using `empirical' phase correction coefficients}
\label{sec:results}
\subsection{Direct use of estimated phase correction coefficients}

The simplest way of using the empirically determined phase correction
coefficients is simply to apply them directly to correct the phases
for science observations, possibly interpolating between adjoining
determinations. If the empirical estimate of the coefficients is made
close in time and elevation to the science target, these should
provide good correction.

In practice this direct approach may not be desirable because:
\begin{itemize}
  \item Changes in elevation (possibly even azimuth, if it is a large
    change) between calibration and science observations, or during
    science observation, will cause an inaccuracy in the phase
    correction coefficients
  \item More time than necessary may be spent on calibration
  \item It may be difficult to predict the dispersive part of the
    phase since model parameters will be poorly constrained
\end{itemize}
Beside the elevation change, it is difficult at this time to predict
the limiting factors of this direct approach, but it should be
possible to test this soon as ALMA starts to be commissioned at the
high site.

\subsection{Constraining the model parameters}
\label{sec:constr-model-param}

A more efficient way of using the empirically determined ${\rm
  d}L/{\rm d} T_{{\rm B},i}$ values is to use them as an additional
observational input to the Bayesian analysis procedure described by
\cite{ALMANikolic587}. In this approach, the observed quantities would
be:
\begin{enumerate}
\item The four observed absolute sky brightness temperatures,
  $T^*_{{\rm B},i}$ and their associated error distribution
\item The four empirically determined correlations between path and
  temperature fluctuations, $\left({\rm d}L/{\rm d} T_{{\rm
      B,}i}\right)^{*}$, and the associated error which due to the
  simplified fitting procedure described in
  Section~\ref{sec:linefitting} will be parametrised by standard
  deviation only, $\sigma_{a,i}$
\end{enumerate}
Since an atmospheric model can directly predict both the absolute sky
brightness and the correction coefficients, we can write down the
joint likelihood function for all these observables:
\begin{equation}
\log{L} = - \sum_{i} \left[ \left( \frac{ T^*_{{\rm
       B},i} - T_{{\rm B},i}}{\sigma_{T,i}} \right)^{2} -
\left(
\frac{\left({\rm d}L/{\rm d} T_{{\rm
      B,}i}\right)^{*} - {\rm d}L/{\rm d} T_{{\rm
      B,}i}}{\sigma_{a,i}}
\right)^2\right]
\end{equation}
where the index $i$ runs from 1 to 4, corresponding to the four
channels of the WVRs and I've again assumed that the errors on the
absolute sky brightness are normally distributed.  This likelihood
function, in conjunction with any relevant priors, can then be
analysed using the Markov Chain Monte Carlo (MCMC) procedure as
described in the previous memo.

\begin{figure*}

\begin{tabular}{c}
  \subfloat[100\% error on ${\rm d}L/{\rm d} T_{\rm B}$]{
  \includegraphics[clip,width=0.33\linewidth]{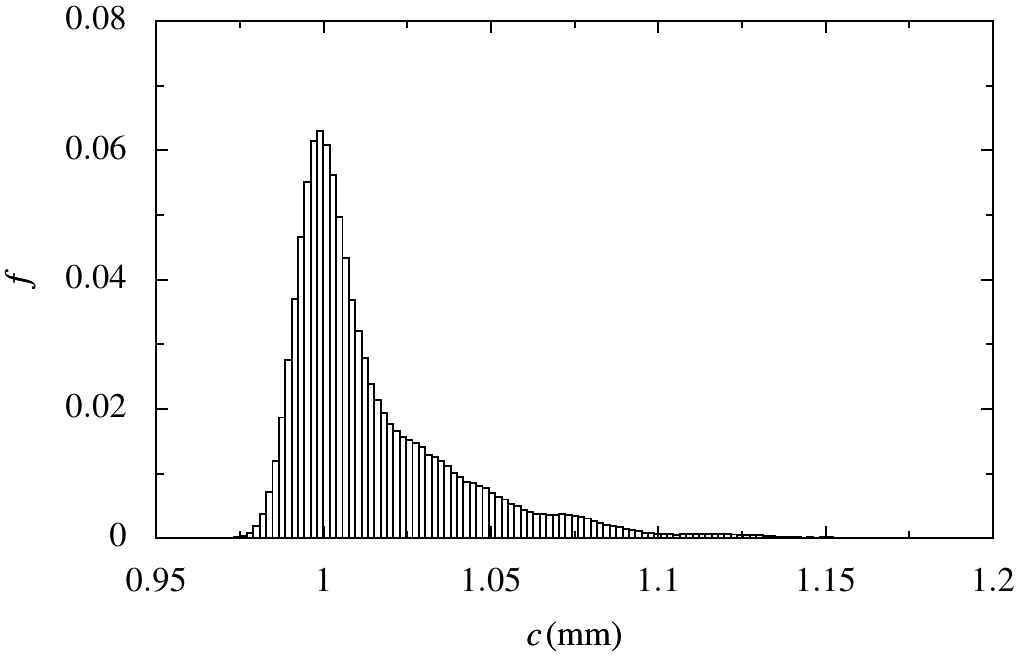}  
  \includegraphics[clip,width=0.33\linewidth]{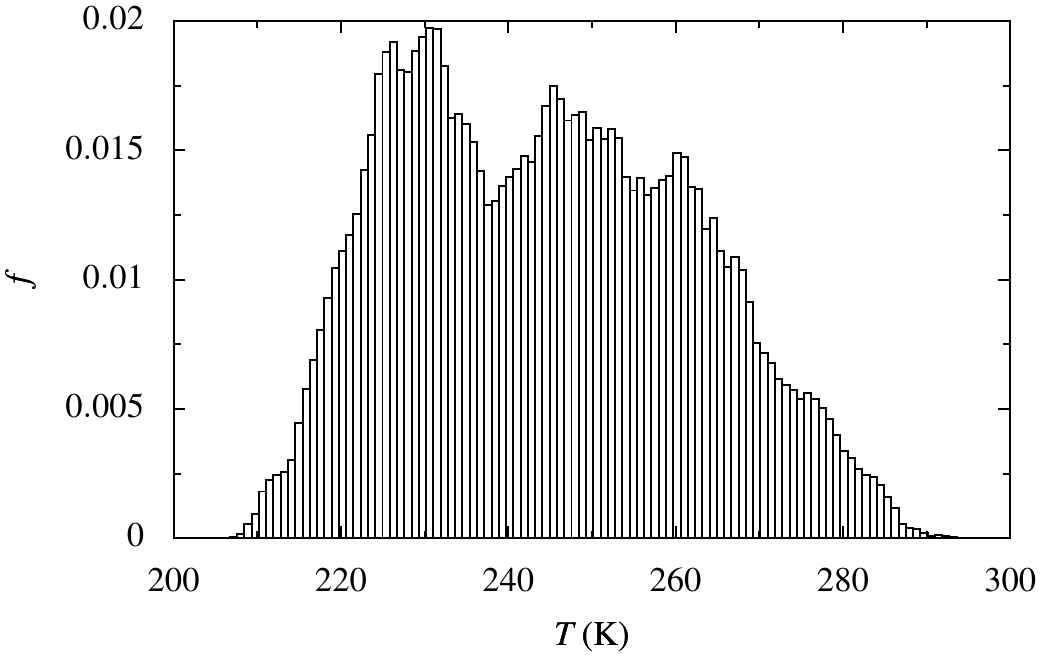}
  \includegraphics[clip,width=0.33\linewidth]{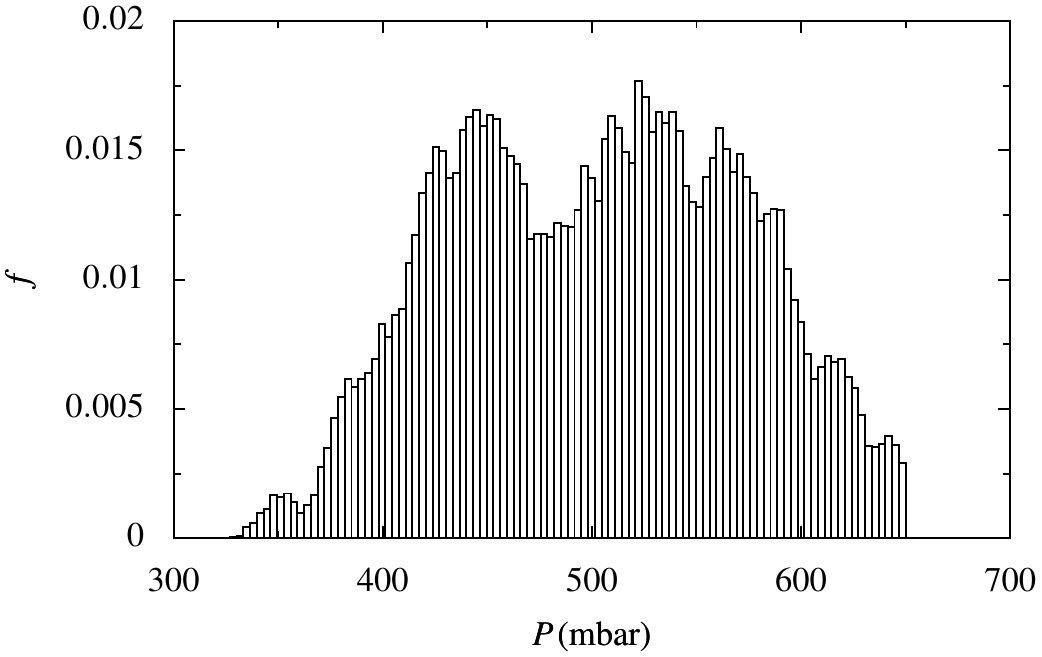}}
\\
  \subfloat[10\% error on ${\rm d}L/{\rm d} T_{\rm B}$]{
  \includegraphics[clip,width=0.33\linewidth]{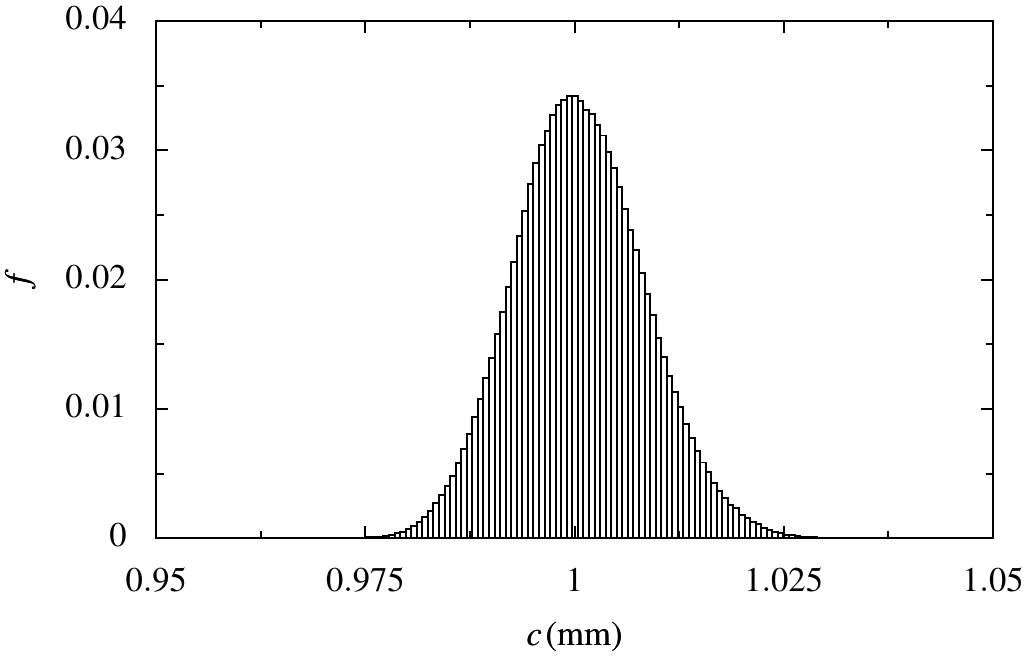} 
  \includegraphics[clip,width=0.33\linewidth]{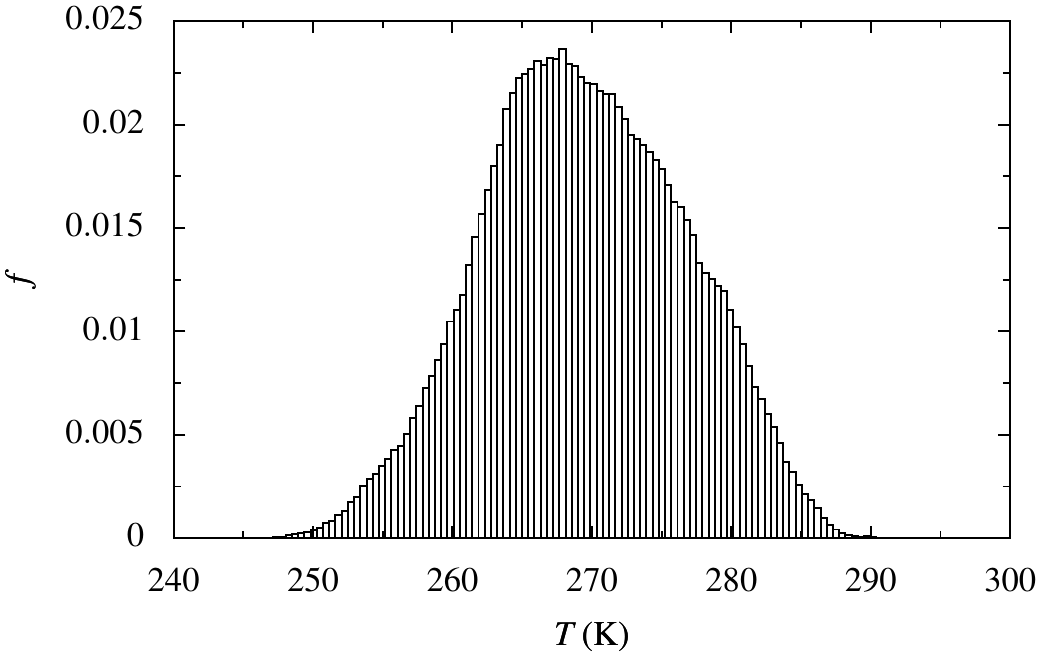}   
  \includegraphics[clip,width=0.33\linewidth]{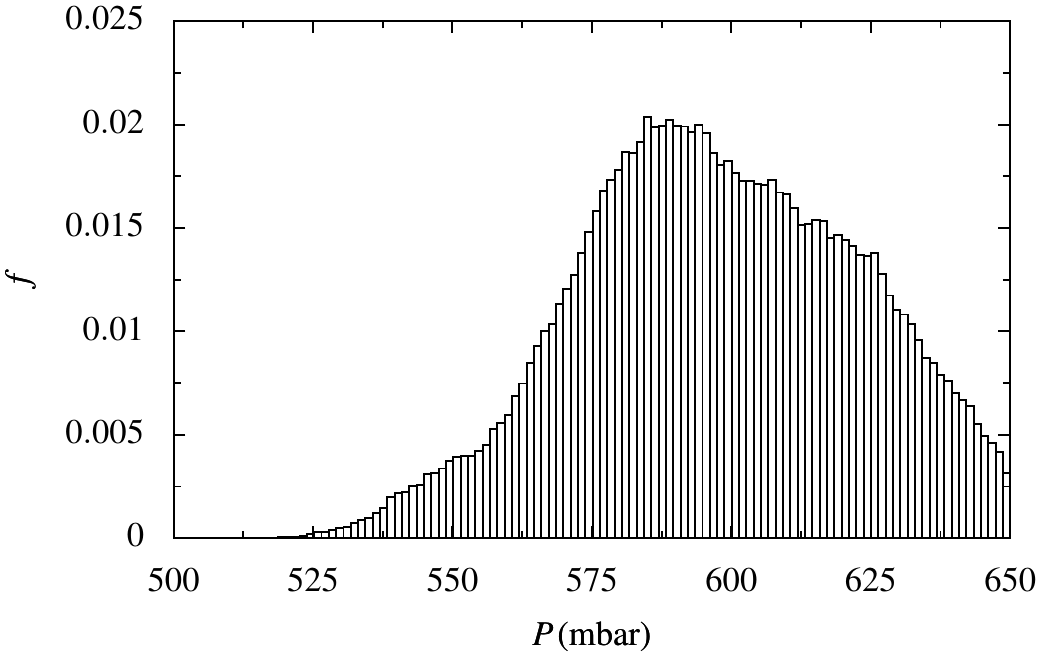}}
  \\
  \subfloat[2\% error on ${\rm d}L/{\rm d} T_{\rm B}$]{
  \includegraphics[clip,width=0.33\linewidth]{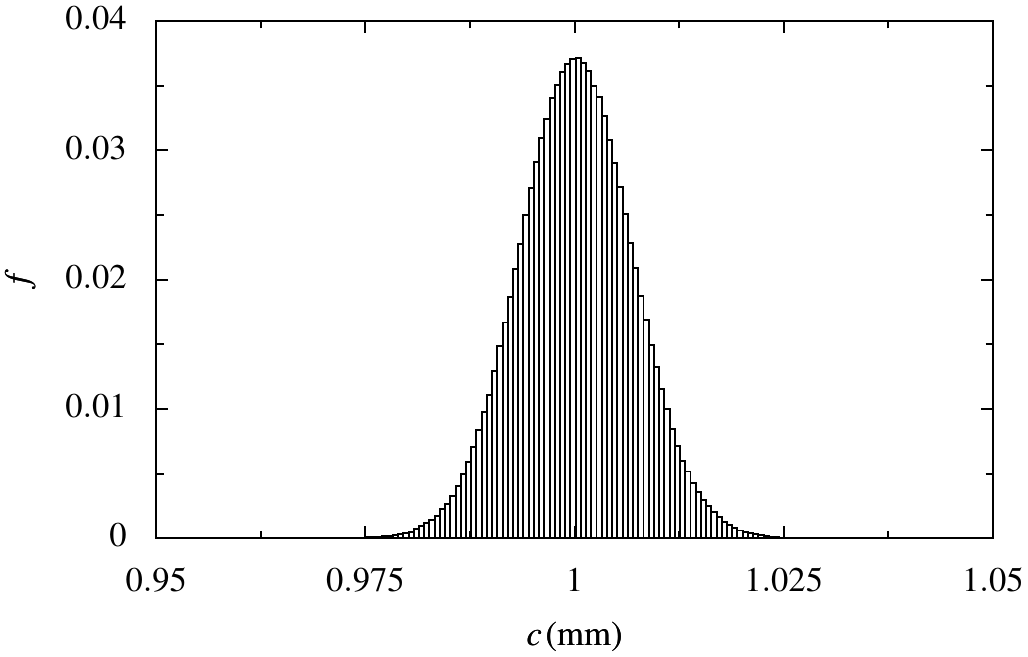} 
  \includegraphics[clip,width=0.33\linewidth]{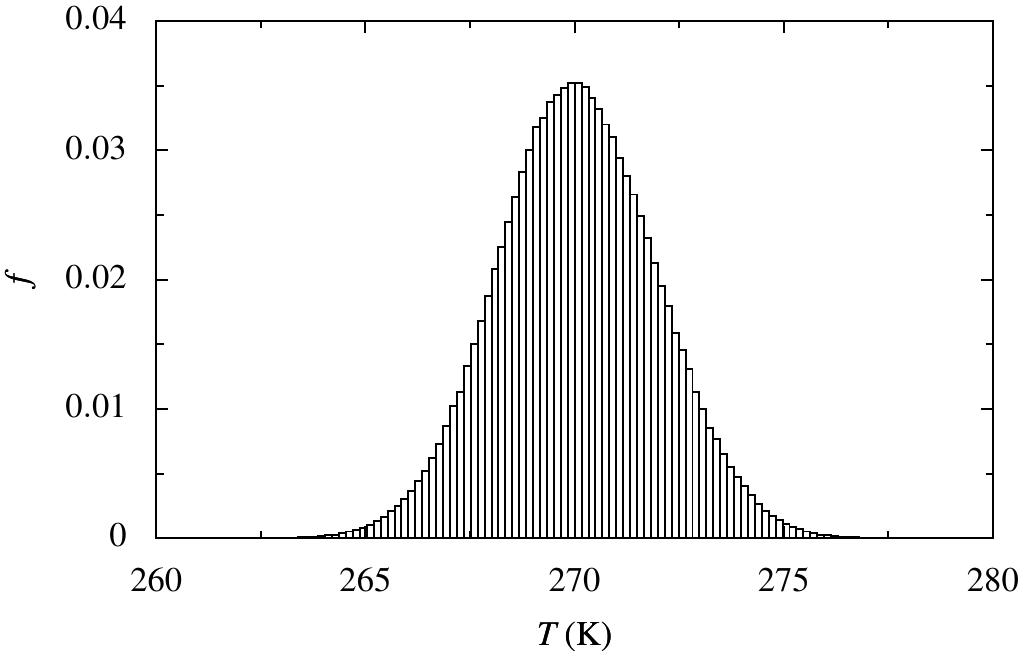}   
  \includegraphics[clip,width=0.33\linewidth]{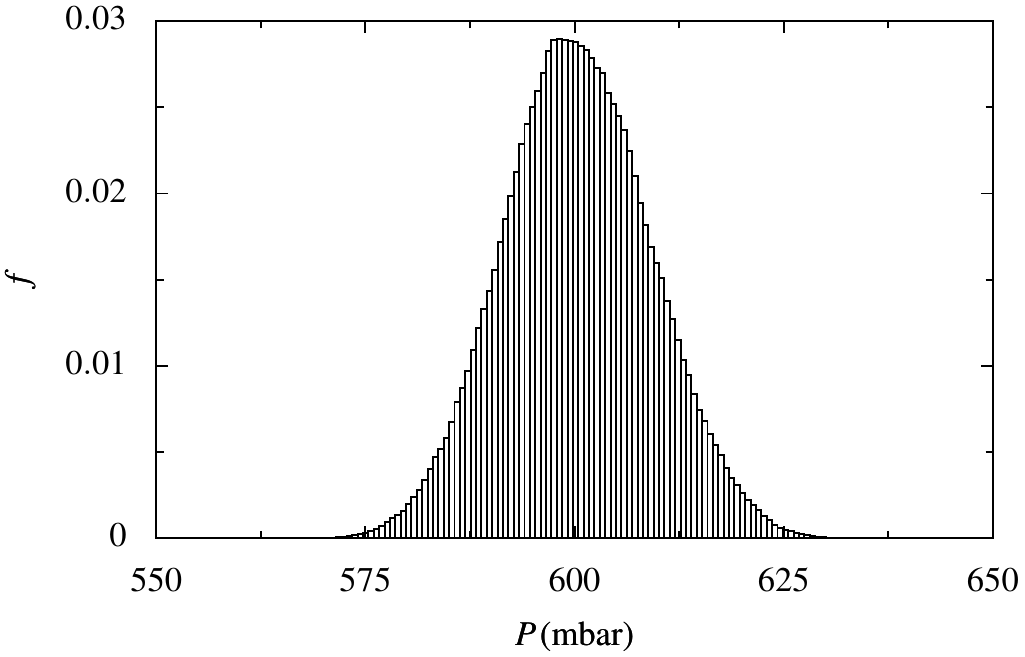}}
  \\
  \subfloat[0.5\% error on ${\rm d}L/{\rm d} T_{\rm B}$]{
  \includegraphics[clip,width=0.33\linewidth]{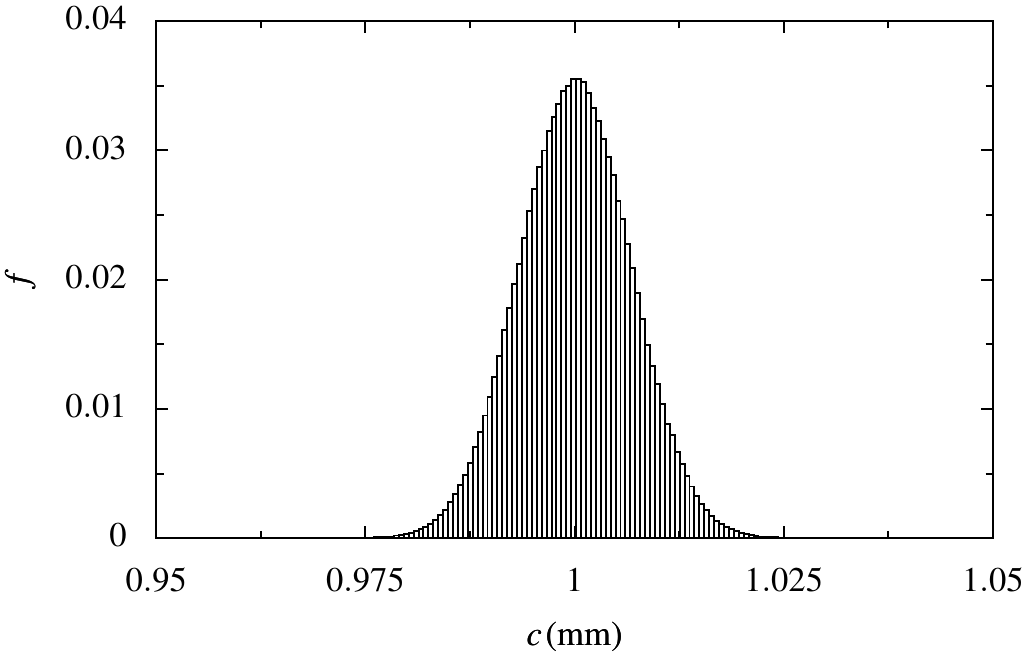}
  \includegraphics[clip,width=0.33\linewidth]{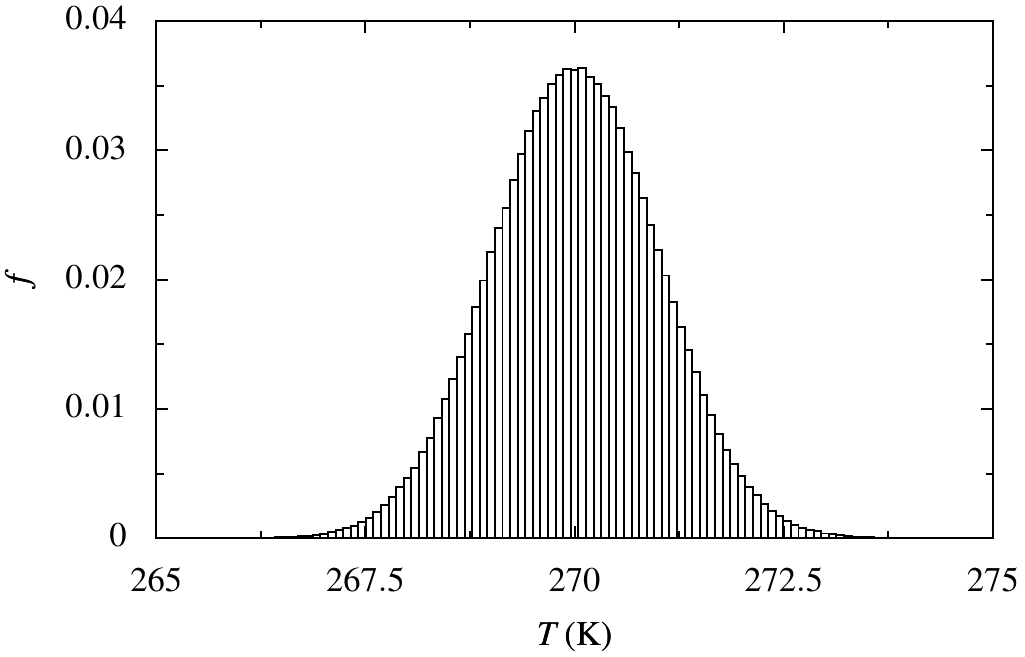}  
  \includegraphics[clip,width=0.33\linewidth]{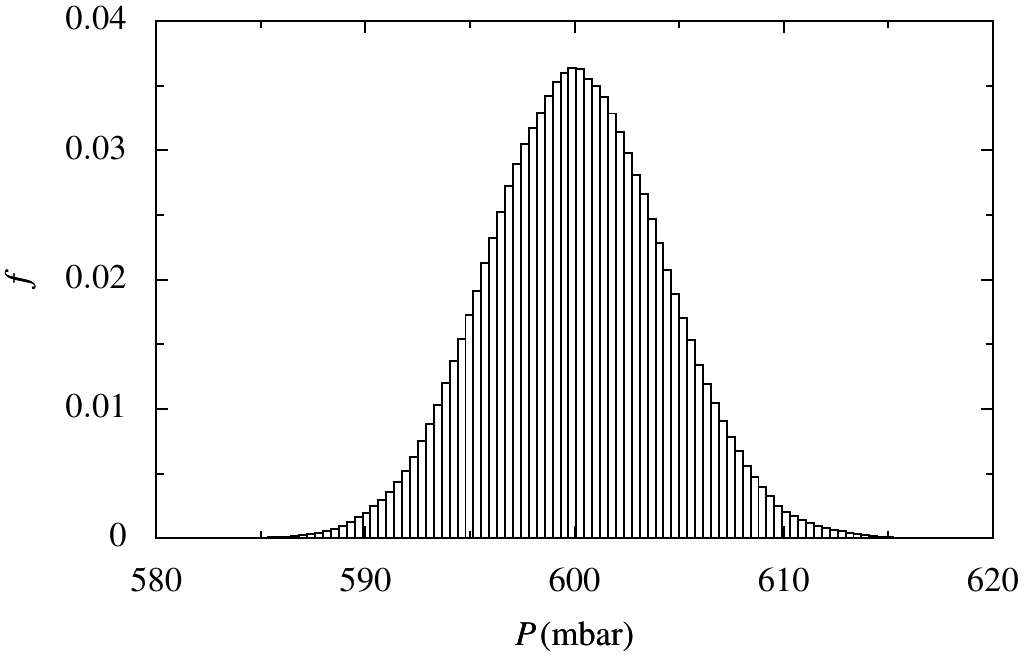}}
\end{tabular}

\caption{Marginalised distributions of model parameters when the
  retrieval is made simultaneously from the absolute brightness
  temperatures (assuming 1\,K Gaussian error on each of the four
  channels) and from the correlation between path and brightness
  fluctuations with Gaussian errors as shown below each of the
  panels. A weak prior on the possible temperature and pressure of the
  water vapour is applied to these retrievals, i.e., the constraint
  that the temperature is between 200\,K and 320\,K and the pressure
  between 100 and 650\,miliBar (this is the row 1 of Table 1 in
  \citealt{ALMANikolic587}).}
\label{fig:modelconstraint}
\end{figure*}

The first question to consider is: how much do the empirical phase
correction coefficients constrain the model parameters in the simple
model of a single, thin, water vapour layer?  As in the previous memo,
this can be analysed using the calculated distributions of model
parameters using a simulated data point as the input. In this case,
the simulated model parameters are $c=1.0\,\unit{mm}$,
$T=270\,\unit{K}$, $P=580\,\unit{mBar}$ which are then used to
calculate both four sky brightness temperatures and the four phase
correction coefficients.

As before, I then assign a 1\,K normally distributed random error on
the absolute sky brightness and a range of errors, from 100\% to
0.5\%, on the phase correction coefficients. The MCMC procedure is the
used to estimate the confidence intervals on the model parameters
$\{c,T,P\}$.

The results of this experiment are shown in
Figure~\ref{fig:modelconstraint}, which shows the inferred
distribution of model parameter for errors on empirical coefficients
which are 100\%, 10\%, 2\% and 0.5\% of their values.  By comparing
the two top rows of this figure, it can be seen that even relatively
noisy empirical estimates of the ${\rm d}L/{\rm d} T_{\rm B}$
significantly constrain the model parameters. 

For example, in the second row of Figure~\ref{fig:modelconstraint},
which corresponds to 10\% error on empirical estimates, the
temperature of the water vapour layer is already constrained to a
range of about 35\,K and the pressure is constrained to a range of
about 100\,mBar.  This is contrast to the top row, which corresponds
to essentially no useful information from the empirical
coefficients. Here the possible ranges of temperature and pressure are
100\,K and 300\,mBar respectively.

It is also useful to look in detail at the effect adding information
from empirical coefficient has on the distribution of the model
parameter representing the water vapour column. Even with the weak
constraints provided by empirical coefficients with 10\% relative
error, it can be seen that the non-Gaussian tail from the distribution
to be eliminated as a possibility. However, tighter constrains from
empirical coefficients with smaller errors \emph{do not\/}
significantly reduce the uncertainty of the retrieved water vapour
column while they moderately improve the errors on the pressure and
temperature parameters.

\begin{figure*}
  \begin{tabular}{cc}
    100\% error on ${\rm d}L/{\rm d} T_{\rm B}$ &
    10\% error on ${\rm d}L/{\rm d} T_{\rm B}$\\
    \includegraphics[clip,width=0.45\linewidth]{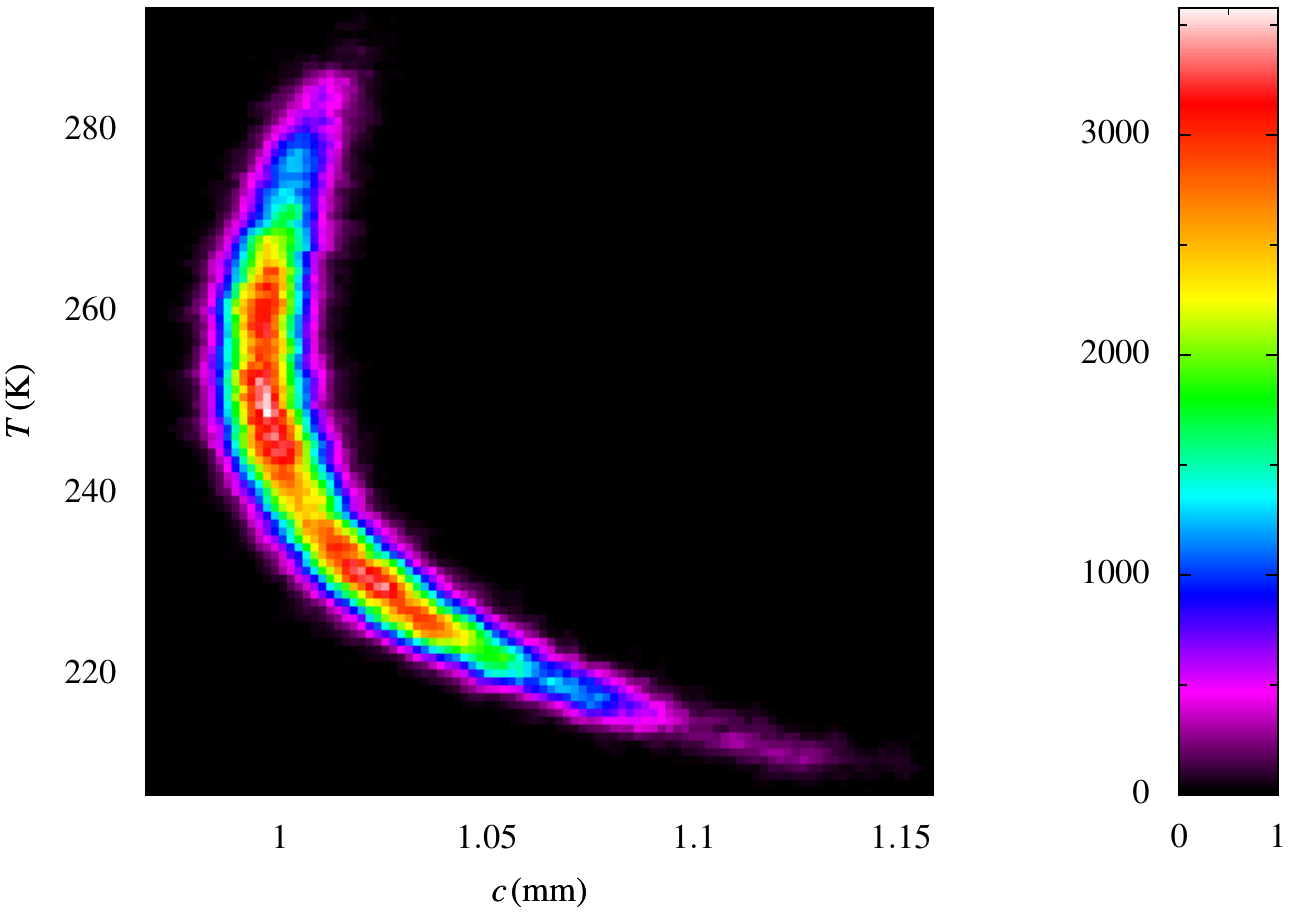}
    &
    \includegraphics[clip,width=0.45\linewidth]{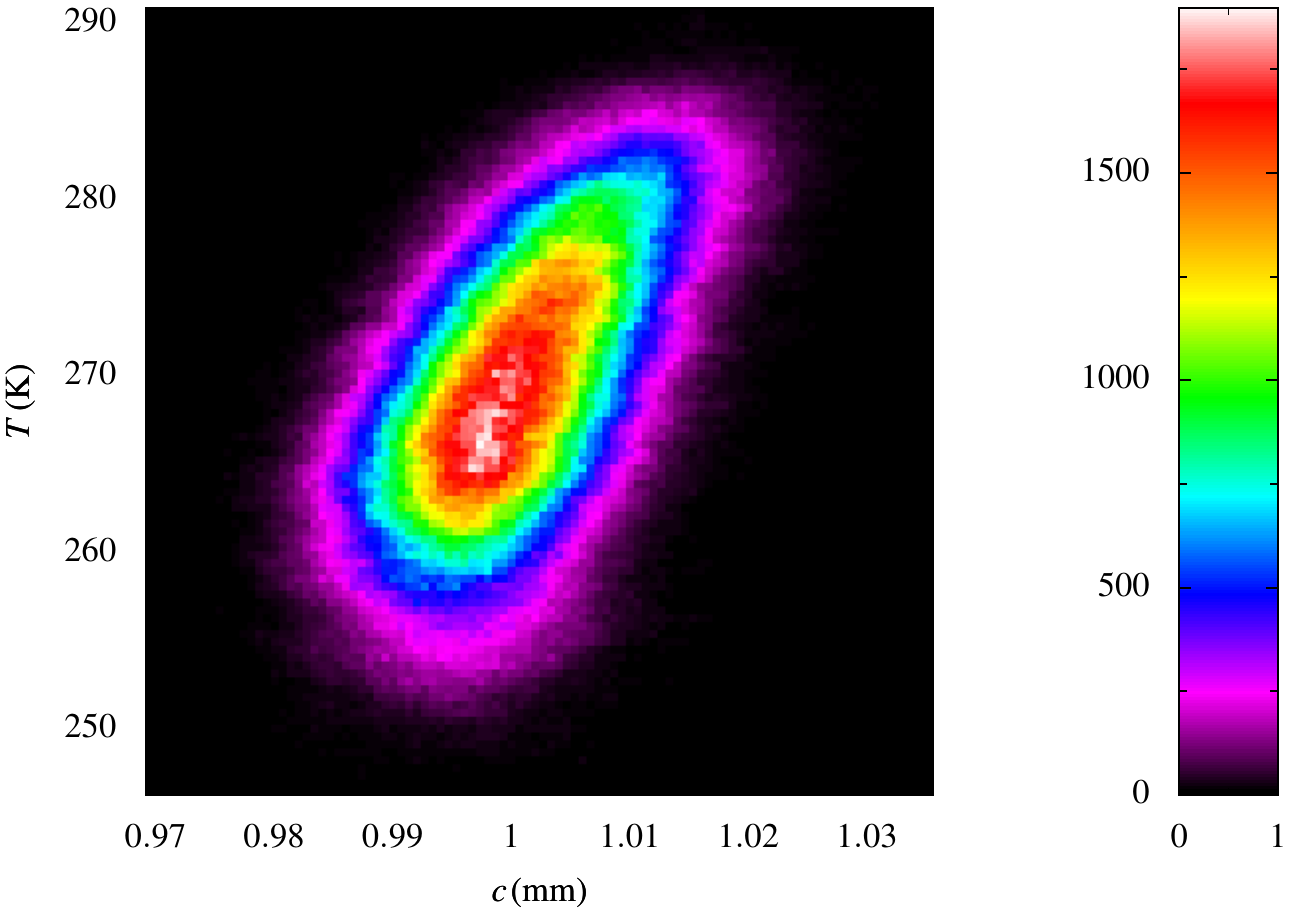}
    \\
    2\% error on ${\rm d}L/{\rm d} T_{\rm B}$ &
    0.5\% error on ${\rm d}L/{\rm d} T_{\rm B}$\\
    \includegraphics[clip,width=0.45\linewidth]{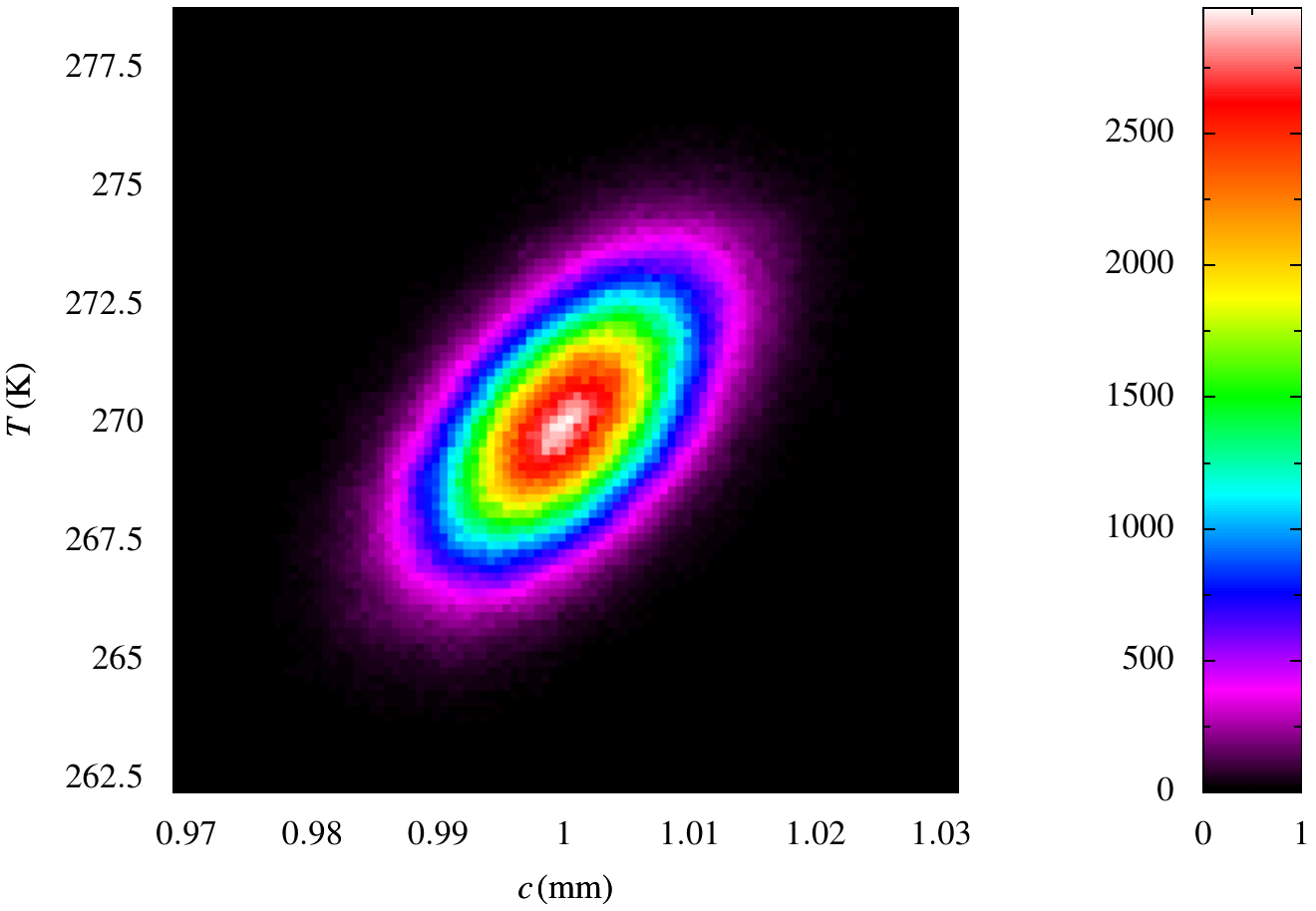}   &
    \includegraphics[clip,width=0.45\linewidth]{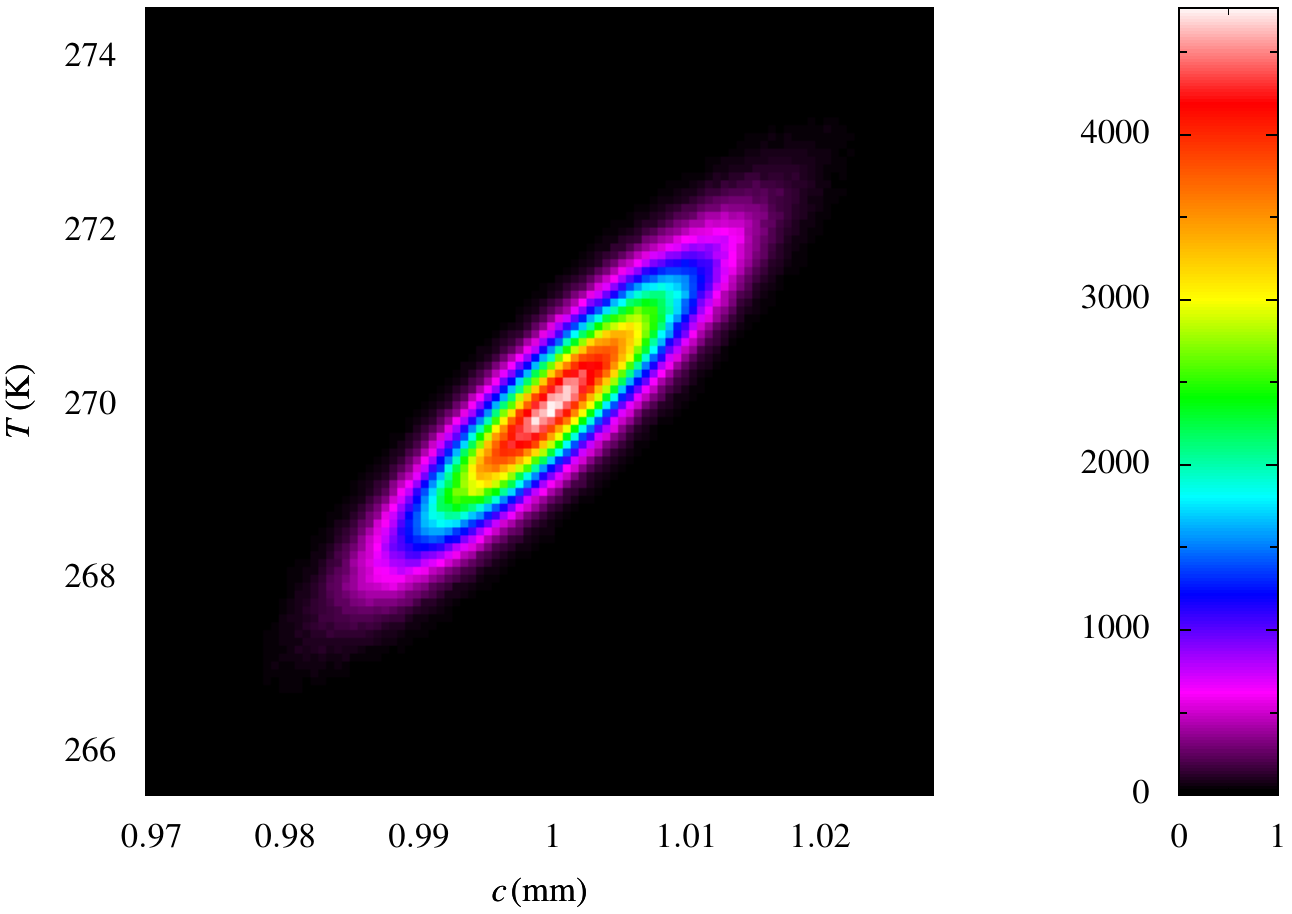}   
  \end{tabular}
  \caption{Joint distribution of the model parameters representing the
    water vapour column (horizontal axis) and the temperature of the
    water vapour layer (vertical axis) based on inference with the
    information from empirical phase correction coefficients included.
    The four panels are for the four different assumed errors on the
    observed coefficients.  Note that the scales changes between the
    panels. }
  \label{fig:joint}

\end{figure*}

The reason for this can be understood from Figure~\ref{fig:joint},
which shows the change in the joint distribution of parameters
representing water vapour column and temperature, as the uncertainty
on empirical coefficients is reduced:
\begin{itemize}
  \item The improvement in the inference of the water vapour column
    between the 100\% and 10\% error on coefficients (i.e., the first
    and second rows) is significant and is in turn due to the better
    constraints on temperature of water vapour which is provided by
    the empirical coefficient measurement
  \item There is little further improvement in the water vapour
    column. This is because this parameter, when there are no
    degeneracies, is already strongly constrained by the absolute
    brightnesses measured by the WVRs
  \item Over a smaller temperature range (in this case $\sim$
    266\,K--275\,K), there is a degeneracy between water vapour and
    the temperature, which prevents strong further constraints on the
    latter
\end{itemize}

Overall, the result illustrates that even poorly estimated empirical
phase correction coefficients can significantly improve the
constraints on the model parameters that are not already well
determined by the absolute measurements of the WVRs, in particular by
removing some of the degeneracies present in the models. The
significance of this increases in more complex physical models with
more parameters, in which further degeneracies are bound to appear.

\begin{figure*}

\begin{tabular}{c}
  \subfloat[100\% error on ${\rm d}L/{\rm d} T_{\rm B}$]{
  \includegraphics[clip,width=0.33\linewidth]{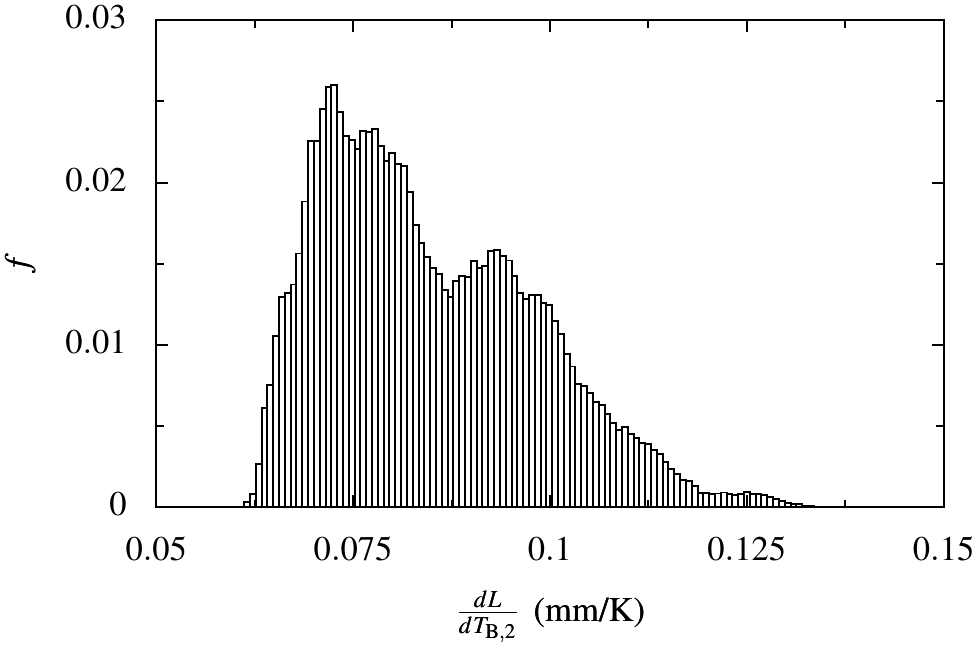}  
  \includegraphics[clip,width=0.33\linewidth]{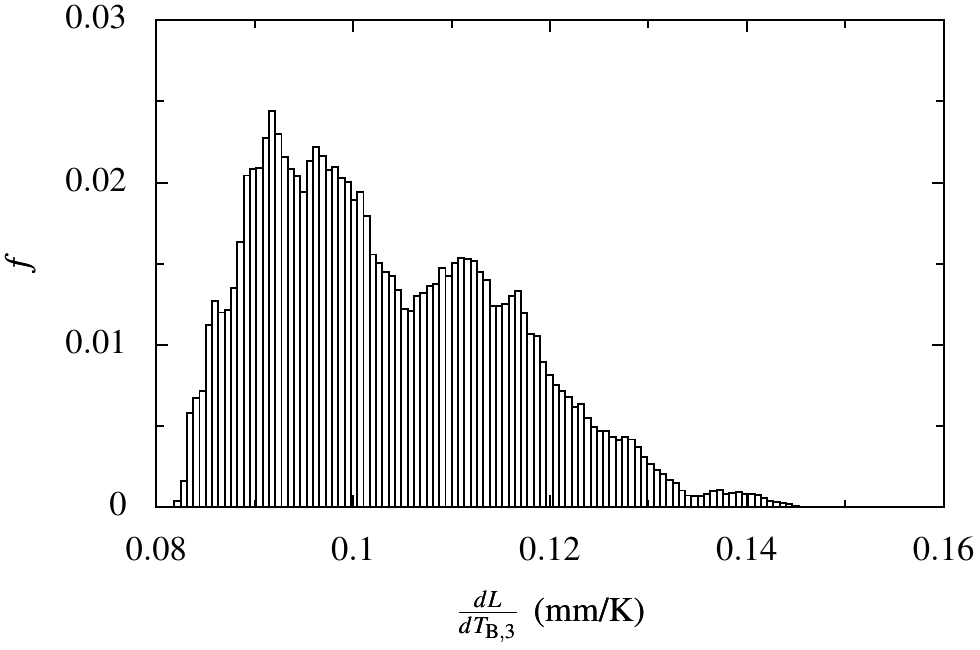}  
  \includegraphics[clip,width=0.33\linewidth]{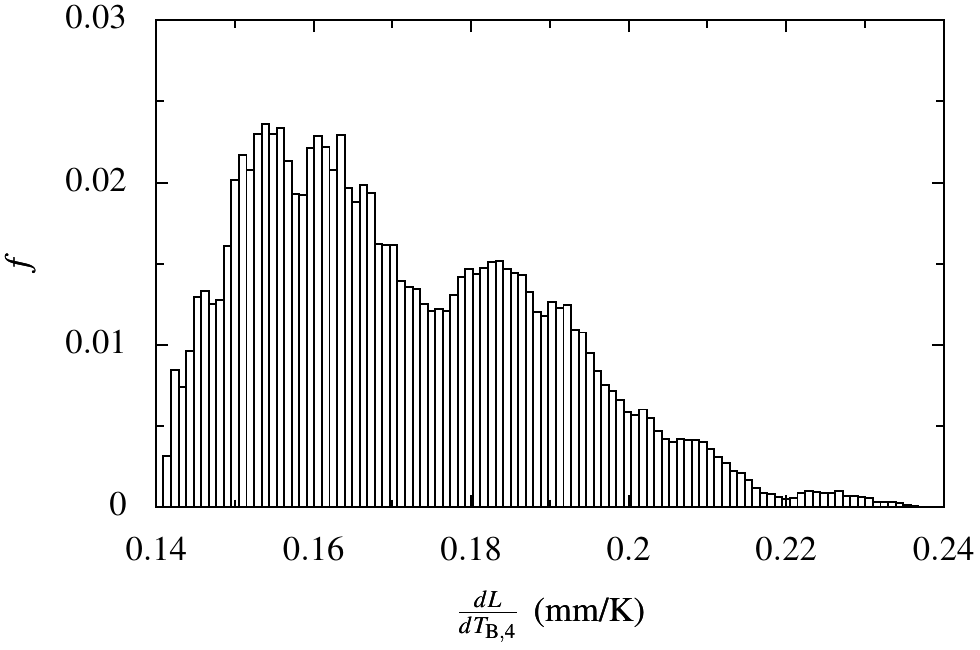}  
  }\\
  \subfloat[10\% error on ${\rm d}L/{\rm d} T_{\rm B}$]{
  \includegraphics[clip,width=0.33\linewidth]{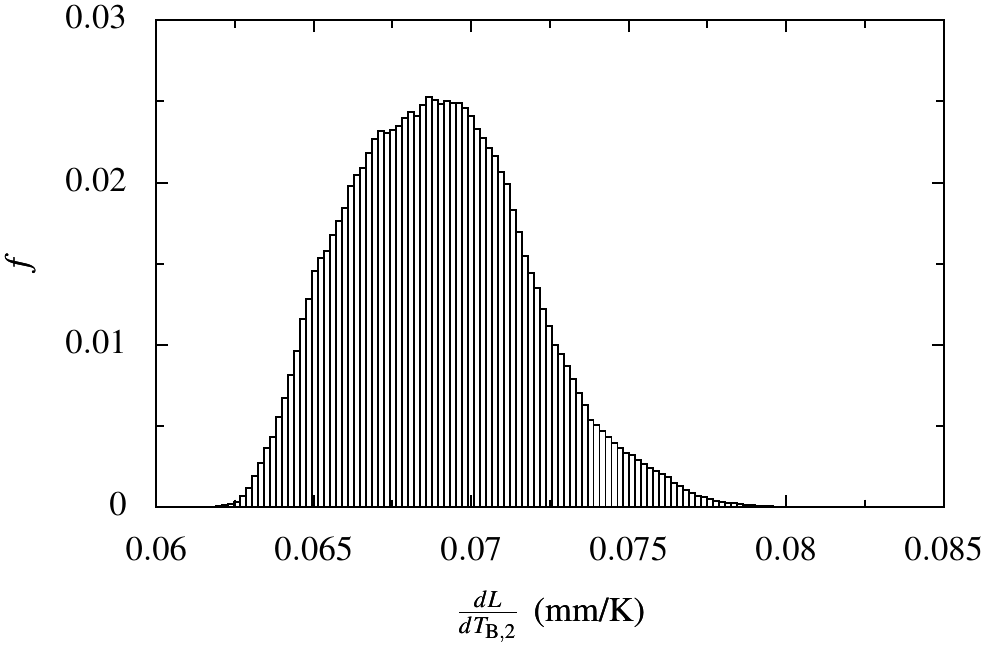}  
  \includegraphics[clip,width=0.33\linewidth]{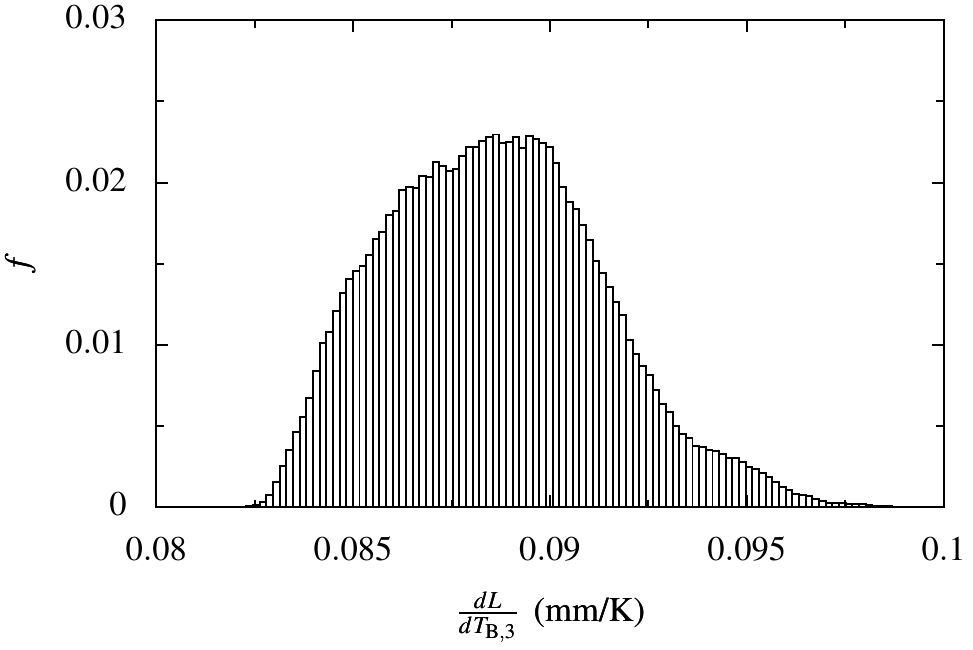}  
  \includegraphics[clip,width=0.33\linewidth]{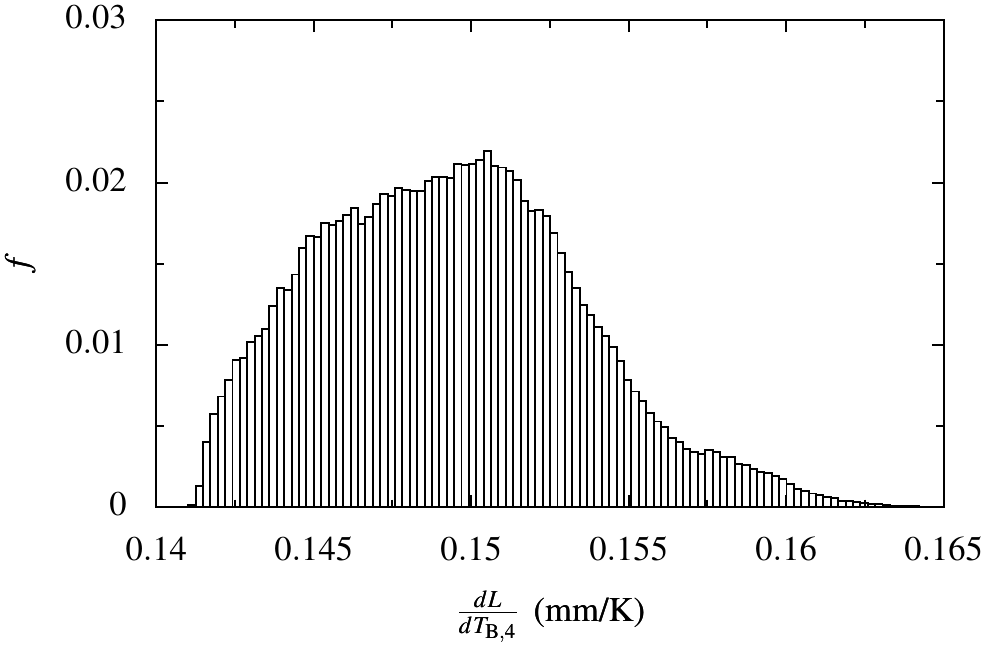}  
  }\\
  \subfloat[2\% error on ${\rm d}L/{\rm d} T_{\rm B}$]{
  \includegraphics[clip,width=0.33\linewidth]{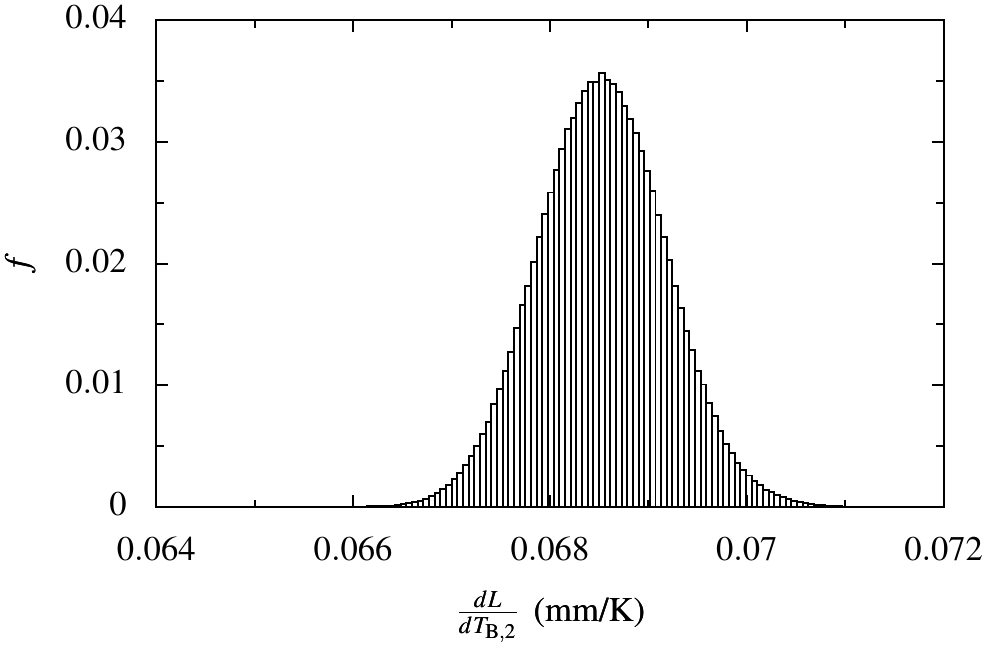}  
  \includegraphics[clip,width=0.33\linewidth]{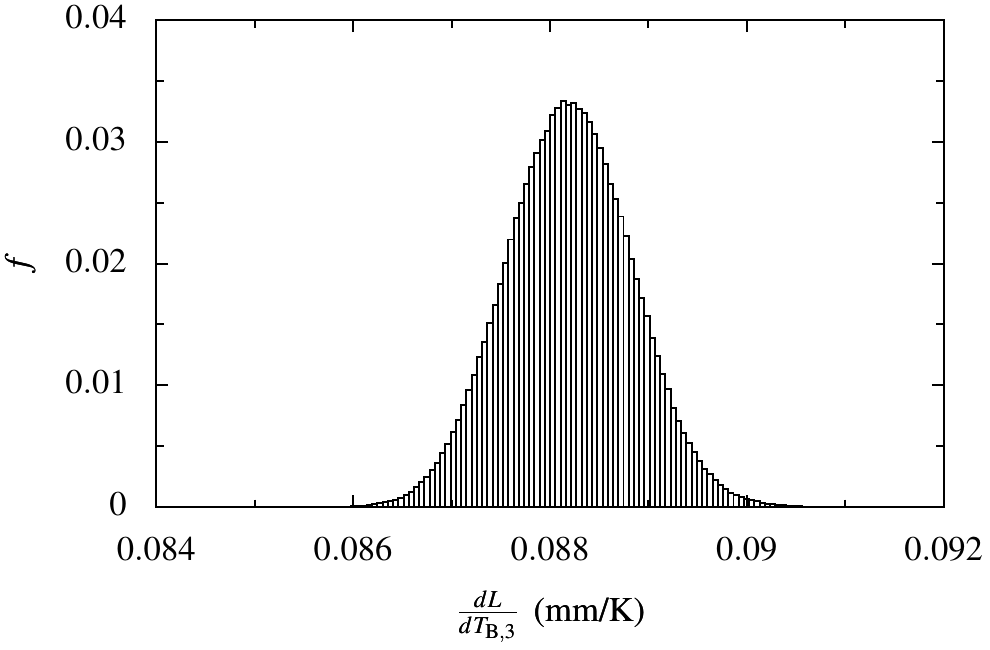}  
  \includegraphics[clip,width=0.33\linewidth]{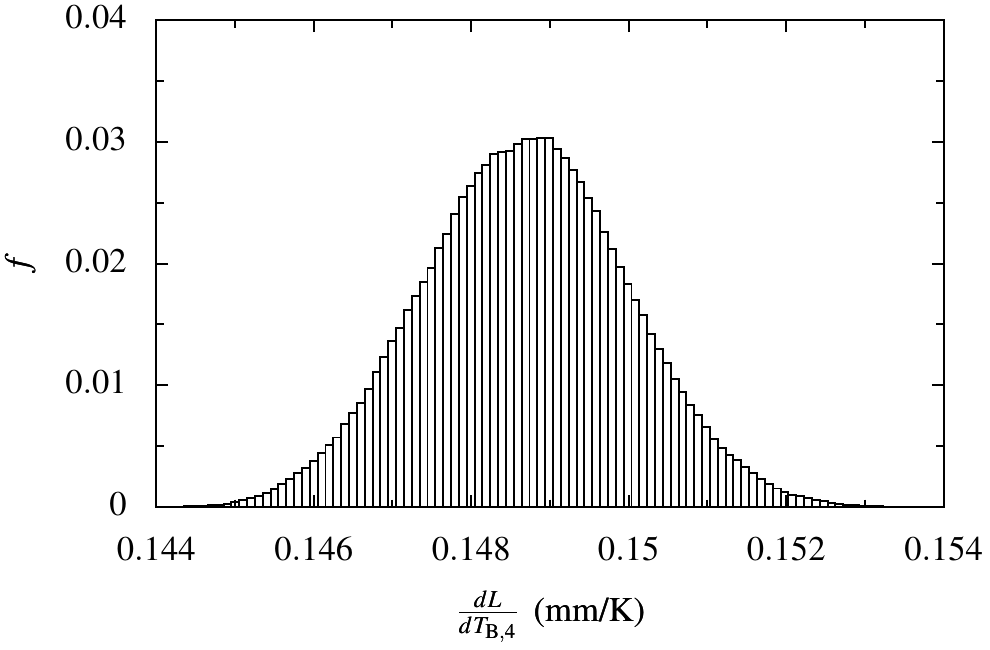}  
  }\\
  \subfloat[0.5\% error on ${\rm d}L/{\rm d} T_{\rm B}$]{
  \includegraphics[clip,width=0.33\linewidth]{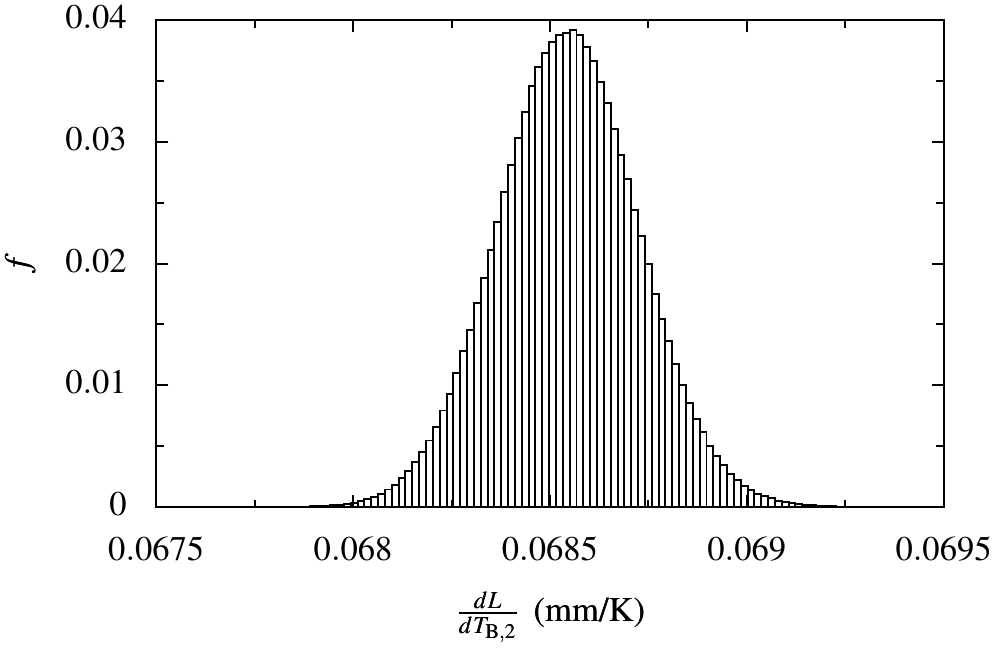}  
  \includegraphics[clip,width=0.33\linewidth]{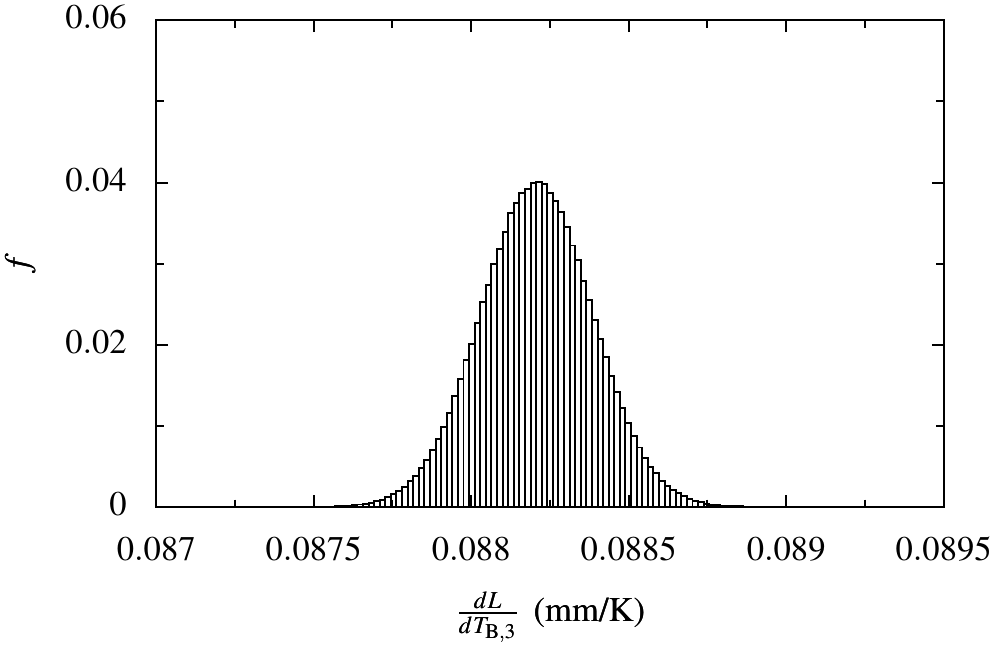}  
  \includegraphics[clip,width=0.33\linewidth]{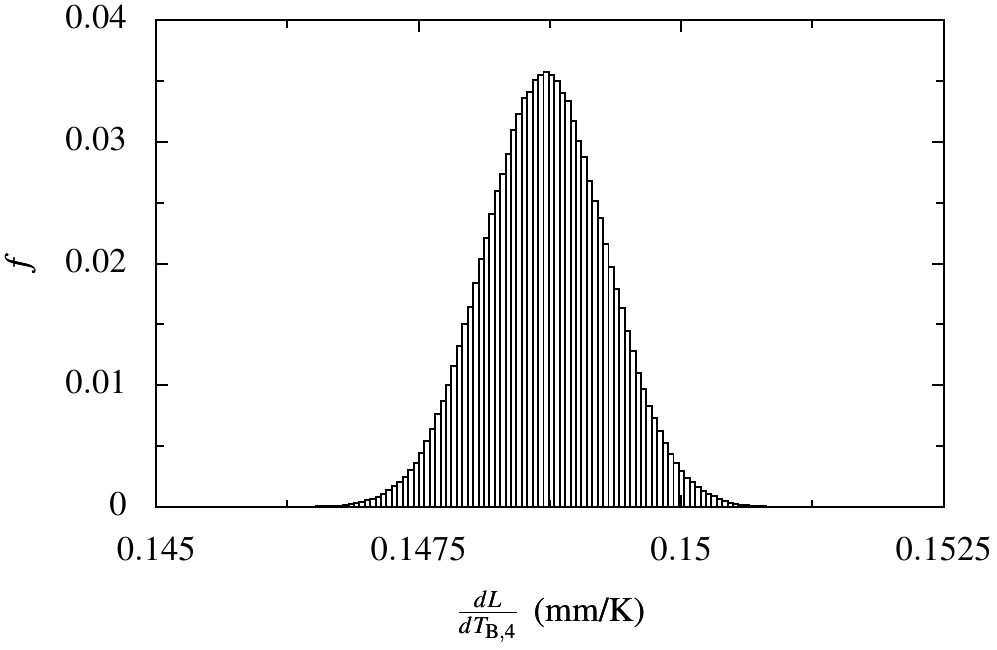}  
  }
\end{tabular}

\caption{Marginalised distributions of the phase correction
  coefficients corresponding to the parameter distributions in
  Figure~\ref{fig:modelconstraint} (to make the panels more readable,
  only three of four coefficients are shown). }
\label{fig:margcoefficients}
\end{figure*}

The final figure in this section (Figure~\ref{fig:margcoefficients})
shows the marginalised distributions of the phase correction
coefficients corresponding to the model parameter distributions shown
in Figures~\ref{fig:modelconstraint} and \ref{fig:joint}. Since
observations of the empirical parameters are used in the inference of
the model parameters, it can be expected that when the errors on the
empirical coefficients are small enough, they completely dominate the
marginalised distribution too. This can indeed be seen for errors of
0.5\% and 2\% -- the distributions are close to the Gaussian
distribution of empirical coefficients which were put into the problem
as observations. On the other hand when errors on empirical
coefficients are of the order of 10\%, the constraints from the
absolute brightness and the empirical coefficients combine to produce
an error on phase correction coefficients which is \emph{smaller} than
the input 10\%.

\subsection{Transfer to different elevation}

\begin{figure*}

\begin{tabular}{cc}
  
  Zenith & 20 degree zenith-angle\\
  \includegraphics[clip,width=0.45\linewidth]{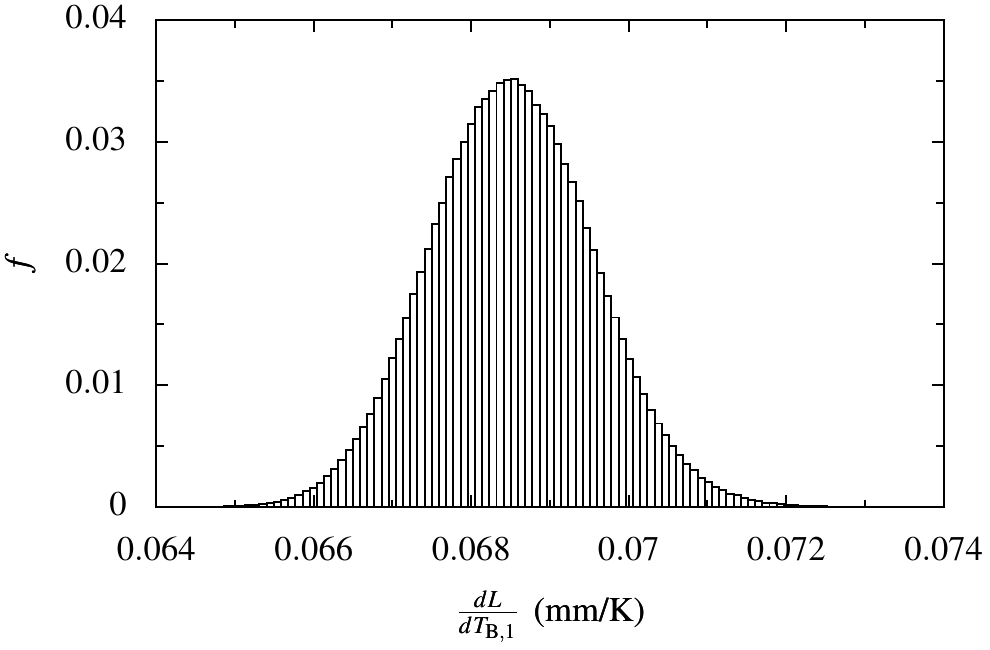}
  &
  \includegraphics[clip,width=0.45\linewidth]{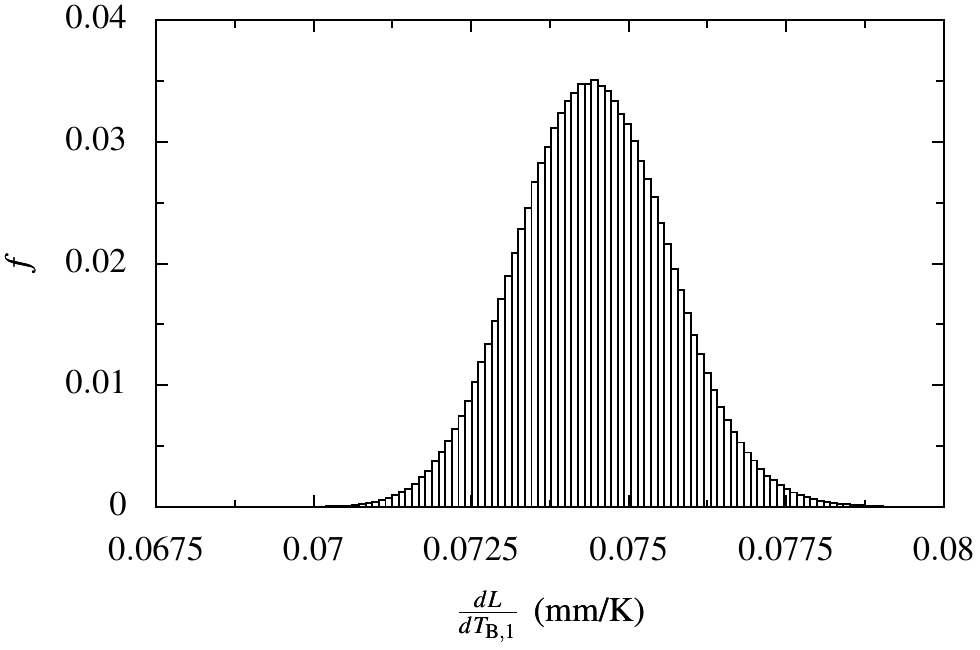}\\
  \includegraphics[clip,width=0.45\linewidth]{figs/pathr/pathweakerr0-02hist-dLdT2}
  &
  \includegraphics[clip,width=0.45\linewidth]{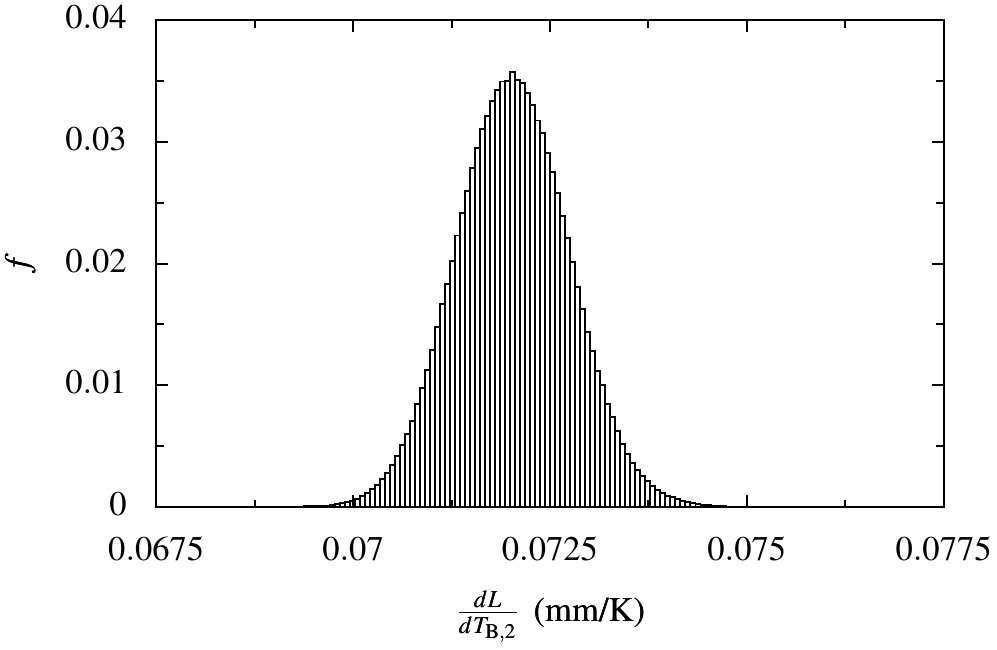}\\
  \includegraphics[clip,width=0.45\linewidth]{figs/pathr/pathweakerr0-02hist-dLdT3}
  &
  \includegraphics[clip,width=0.45\linewidth]{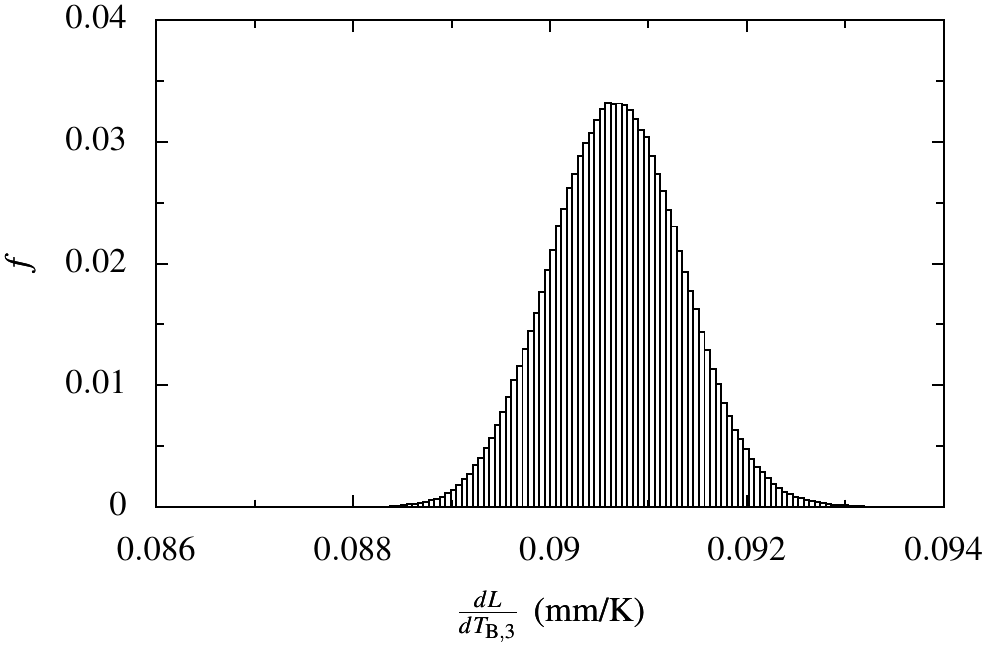}\\
  \includegraphics[clip,width=0.45\linewidth]{figs/pathr/pathweakerr0-02hist-dLdT4}
  &
  \includegraphics[clip,width=0.45\linewidth]{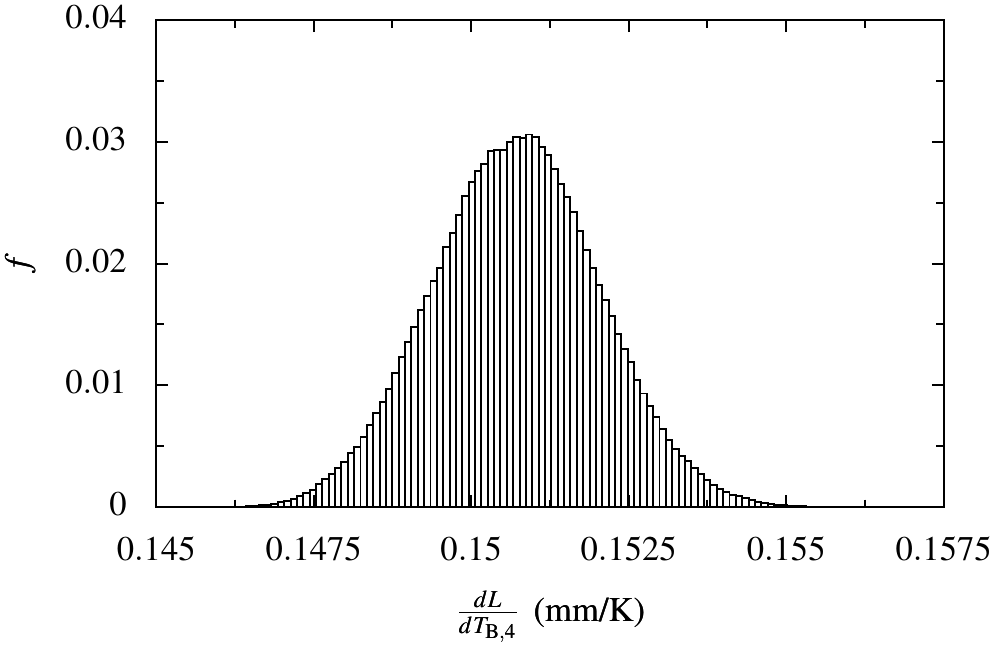}
\end{tabular}
\caption{Transfer of inferred phase coefficients from zenith to
  elevation 20 degrees away.}
\label{fig:transfer}
\end{figure*}

One of the key reasons for combining the information from the
empirical coefficients with a physical model of the atmosphere is that
it provides a way of extrapolating the results to slightly different
conditions.

Probably the most important example of this is a change in the
elevation of the antennas, which will usually be necessary when moving
from the calibration source to the science source and also while
tracking the science source as it rises or sets. By making the usual
approximation of a plane parallel atmosphere we have a direct
prescription on how to modify the parameters of the model. In the case
of the present, very simple, model this is simply to scale the water
vapour column as $\sec(\theta)$ while leaving the other parameters
unchanged. This transformation can be carried out on the full
posterior distribution of the parameters as obtained from the
inference procedure, leading to a new estimate for the distributions
of the correction coefficients at the new elevation.

An example of the procedure of transferring the results of the
inference to a different elevation are shown in
Figure~\ref{fig:transfer}. In this example a simulated observation at
zenith with an assumed 2\% error on empirical coefficients is analysed
to provide the posterior distribution of the model parameters. These
are then transferred to an elevation of 70 degrees (zenith angle of 20
degrees) and the new distribution of phase correction coefficients are
computed. As can be seen the distributions of the coefficients are
shifted between the two elevations corresponding to different levels
of saturation of the 183\,GHz water vapour line. The overall shapes
are still largely the same however, and still approximately normally
distributed, and there are no undesirable tails. This of course is
simply the consequence of the good constraints on temperature and
pressure we obtained at the zenith which we know (in this model, and
probably in practice) will not change with elevation of the antennas.

More advanced extrapolations are also possible.  In particular, the
sky toward the science source will have an intrinsically different
amount of water vapour compared to the calibration source direction,
over and above what is predicted by the plane parallel atmospheric
model and the change in elevation between the two sources. That
intrinsic change can be constrained by using the absolute sky
brightness measurements by the WVRs in the direction of the science
source as an additional input to the retrieval.

A possible implementation of this approach is to have two variables
representing water vapour column, one in the direction of the
calibration source at time of empirical coefficients were determined
and one in the direction of the science target at the current
time. The two variables can then be tied together with a prior
obtained through experience of how quickly the conditions at the site
change. The observables in this case are the absolute sky brightness
and empirical coefficients in the direction of the calibrator and the
current absolute sky brightness in the direction of the science
target. Such an approach would be able to maximise the time during
which the information obtained from empirical coefficients is useful.

\section{Illustration with data from the SMA}
\label{sec:smaillustration}

In this Section I make use of an one-hour long stretch of test data
obtained with the prototype ALMA water-vapour radiometers at the
Submillimetre Array (SMA) to illustrate the above concepts of
obtaining the empirical phase correction coefficients and using them
in the retrieval of atmospheric parameters. This data set is the same
data set as used in the previous memo of this series
\citep{ALMANikolic587}.

\begin{figure*}

\begin{tabular}{cc}
  \subfloat[Every integration, i.e., every 2.5\,s]{
  \includegraphics[clip,width=0.49\linewidth]{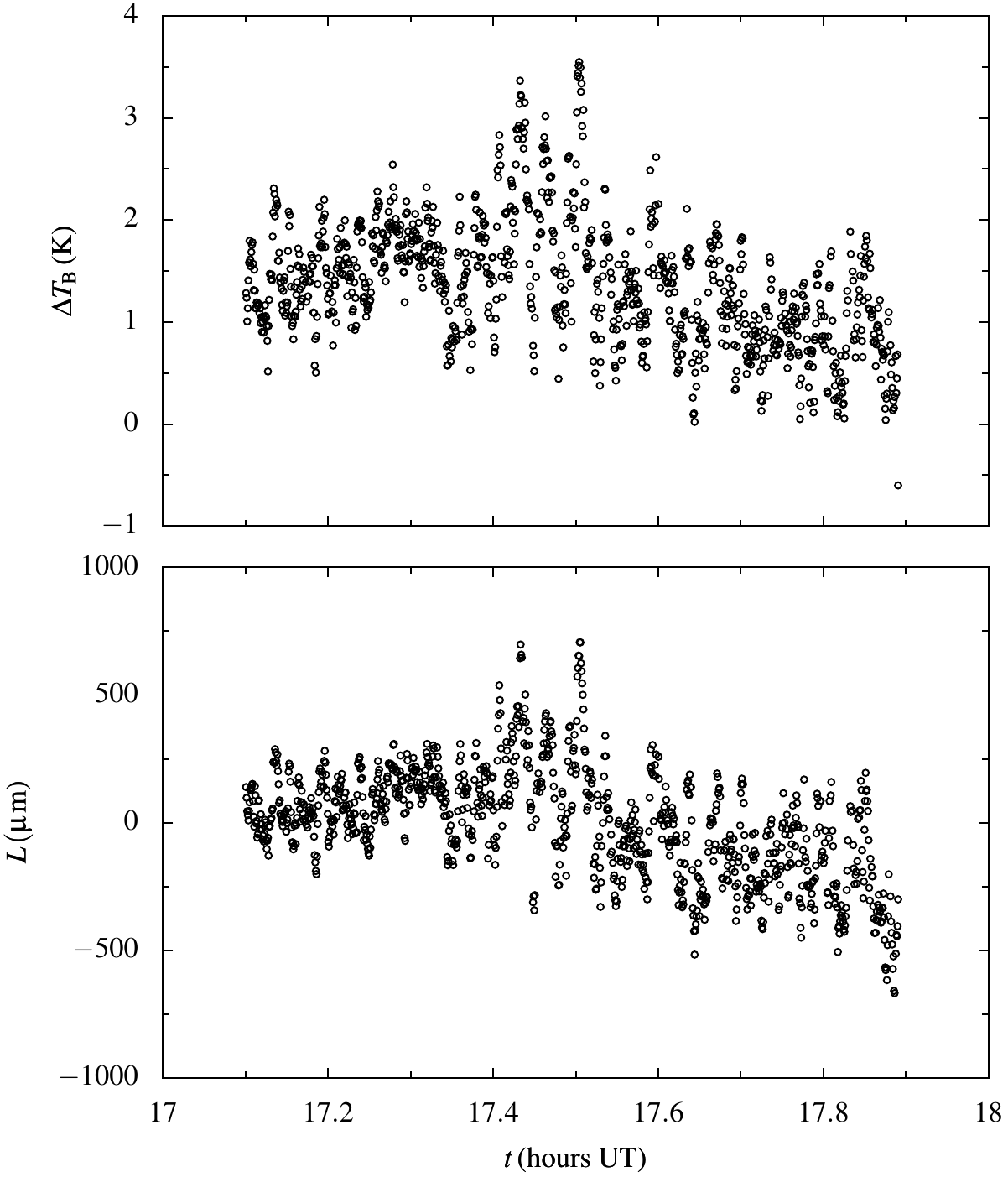}}
  &
  \subfloat[Every 10th integration, i.e., every 25\,s]{
  \includegraphics[clip,width=0.49\linewidth]{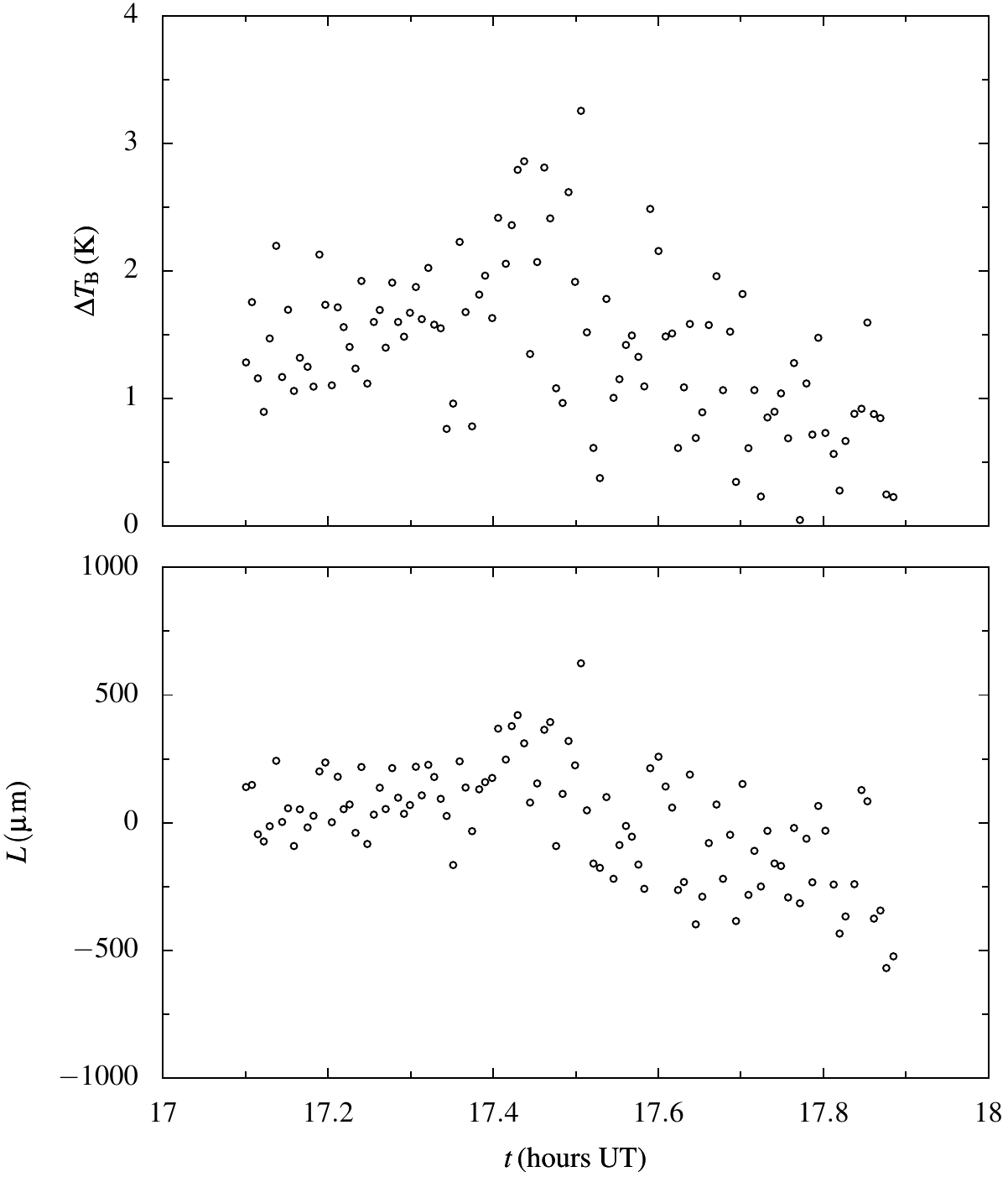}}
  \\
  \subfloat[Every 24th integration, i.e., every 1\,minute]{\includegraphics[clip,width=0.49\linewidth]{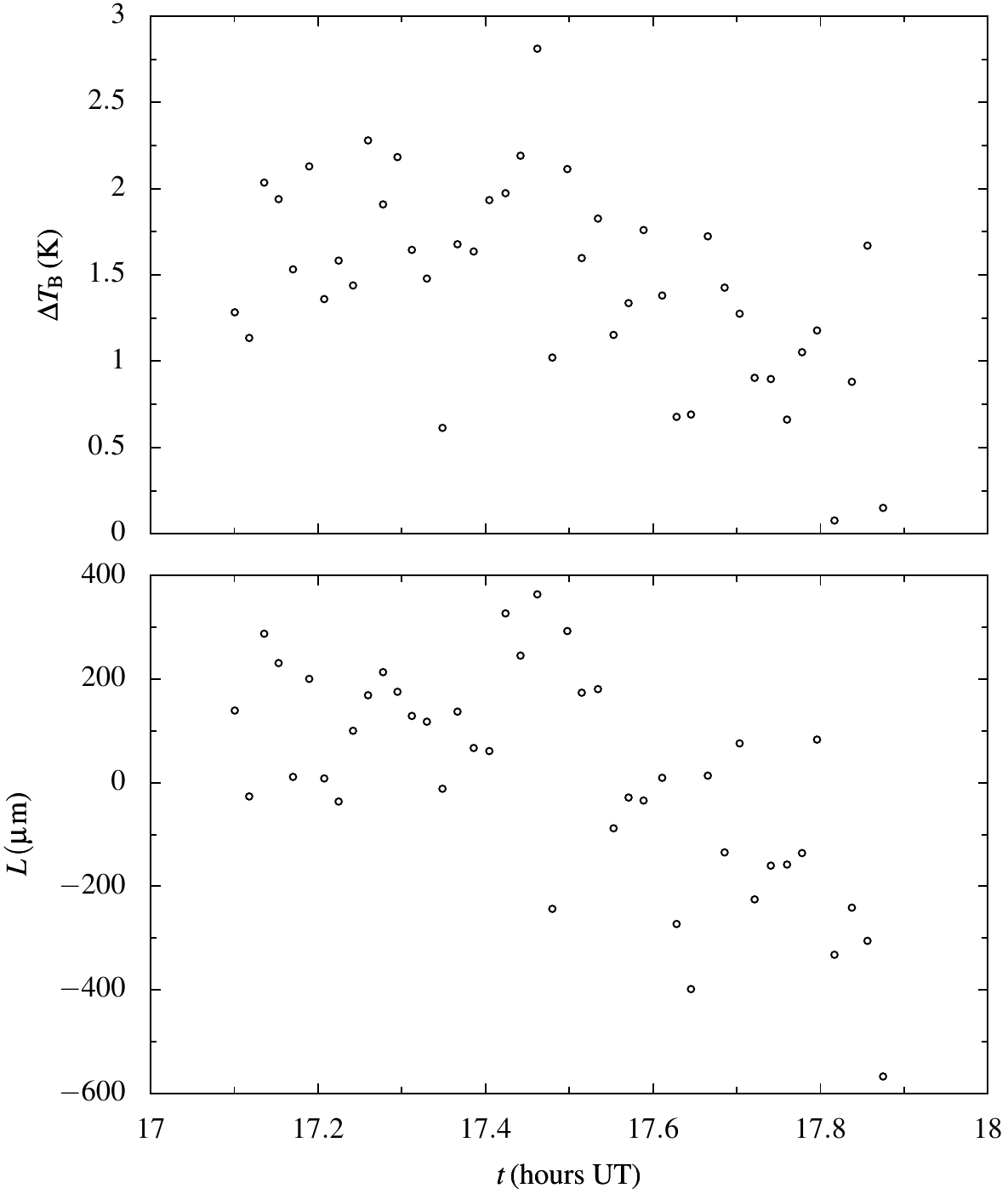}}
  &
  \subfloat[Every 50th integration, i.e., every
    2\,minutes]{\includegraphics[clip,width=0.49\linewidth]{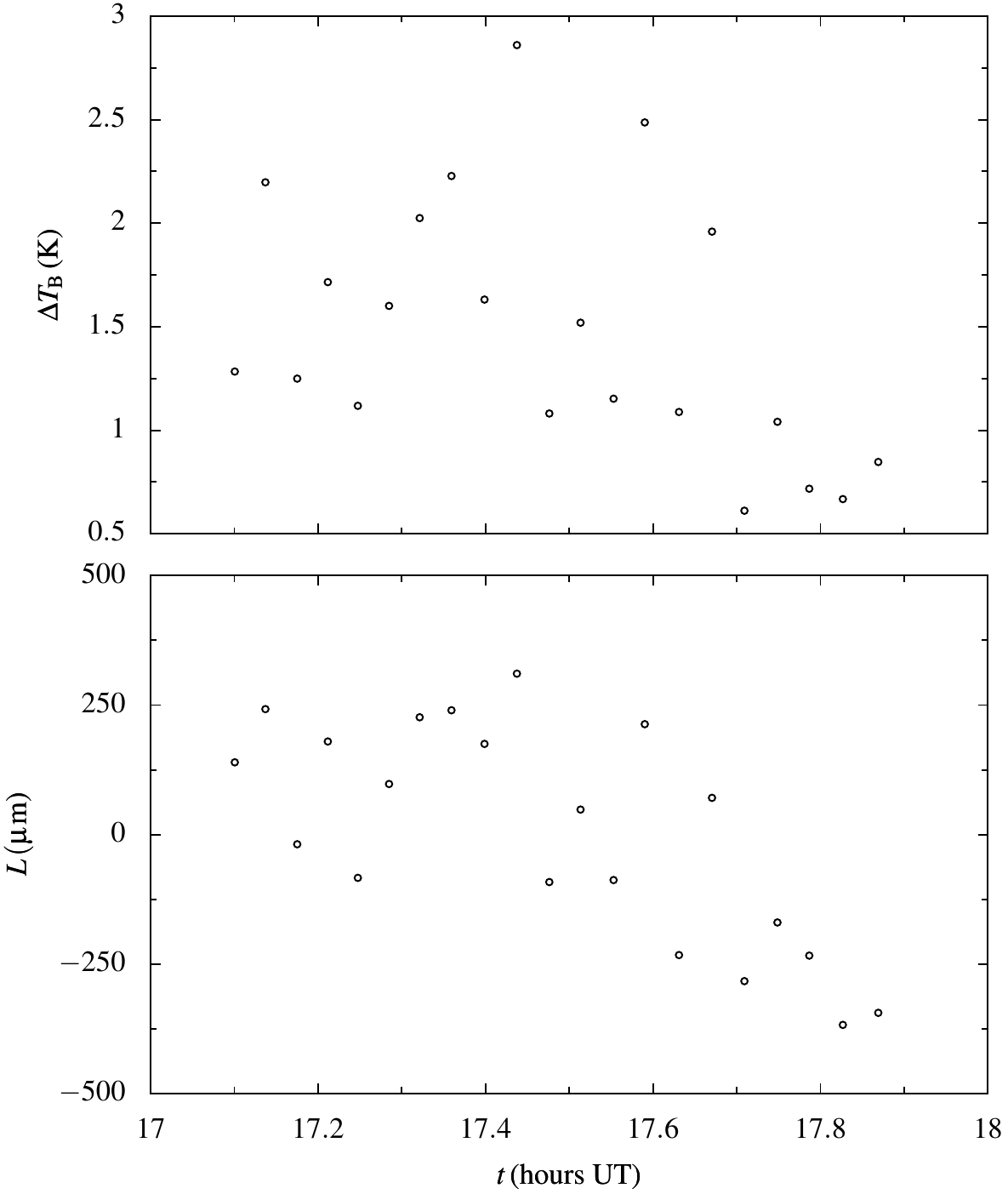}}
\end{tabular}

\caption{Illustration of decimation of single-baseline data from the
  SMA to simulate inference of correlation between sky brightness and
  path fluctuation using fast-switching calibration observations. In
  each panel, the upper plot shows the difference in radiometric
  measurements ($\Delta T_{\rm B}$, using channel four of the WVRs)
  and the bottom plot shows inferred path between the antennas
  ($L$). }
\label{fig:smadata}
\end{figure*}

The original data obtained at the SMA are shown in the top-left panel
of Figure~\ref{fig:smadata}. The top plot in this panel is the
difference between the sky brightness temperatures recorded at each
integration time by the WVRs on the two antennas used in this
test. Only the data from the outermost channel of the WVR is shown (as
discussed in the previous memo, the data from the innermost channels
are not very useful for correction in this case as the water vapour
line is almost optically thick).  The bottom plot is the
interferometric phase recorded between the antennas and converted to a
path-length fluctuation. Like most of the data that was collected at
the SMA, the integration time in this experiment was 2.5\,s.

Also shown in Figure~\ref{fig:smadata} are the trivial simulations of
the type of phase measurement that may be obtained by fast-switching
observations with ALMA. These were made by simply decimating the data
collected at the SMA so that the remaining data resemble
fast-switching observations with cycles of 25, 60 and 120 seconds. It
is clear from these plots that even with a 120 second cycle time there
are correlations between the measured phase and sky temperature
difference which contain useful information about the empirical phase
correction coefficients.

\begin{table}

\begin{tabularx}{\columnwidth}{Xccccc}
\toprule
Interval & Ch 1 & Ch 2 & Ch 3 & Ch 4\\
(s)      & (\micron/K) &(\micron/K) & (\micron/K) & (\micron/K)\\\midrule
2.5      & -161     & 94     & 388 (0.02)  & 351 (0.01)\\
25       & -181     & 88     & 397 (0.22)  & 345 (0.14)\\
60       & -199     & 7      & 423 (0.8)  & 361 (0.44)\\
125      & -275     & 34     & 394 (1.2)  & 335 (0.7) \\\bottomrule      
\end{tabularx}

\caption{Summary of inferred linear coefficient between phase and sky
  brightness fluctuations for the full (top row) and the decimated
  data from the SMA. The values for channels 1 and 2 were calculated
  using the standard least-squares straight line fitting since the
  correlation is so poor that the algorithm of
  Section~\ref{sec:linefitting} produces results with extremely large
  formal errors.  The number in the parenthesis for channels 3 and 4
  is the estimated formal error from the line fitting procedure. }
\label{tab:smal-empirical-coeffs}
\end{table}

By again making the linear assumption
(Section~\ref{sec:linearisation}), these data can be used to estimate
the empirical phase correction coefficients. For these data (as
discussed in the previous memo) channels 1 and 2 have a high optical
depth and are not useful for phase correction. For the same reason,
application of the line fitting with errors in both coordinates
(Section~\ref{sec:linefitting}) produces formal errors much larger
than the actual coefficients. Therefore the line fitting with errors
in both coordinates is only applied to channels 3 and 4 while for
channels 1 and 2 the simple traditional fitting with error in one
coordinate only is used. The data from channels 1 and 2 are not
however used in any other way. The results of this analysis is shown
in Table~\ref{tab:smal-empirical-coeffs}.

Concentrating on channels 3 and 4 which are relevant in this case, it
should first be noted that decimation of the data from the full data
to one data point every 125\,seconds \emph{does not\/} drastically
change the value of the empirical phase correction coefficients,
although some variation is seen and the formal errors on the values
increase. This is consistent with the results of the simulations
(Section~\ref{sec:single-baseline-res}) which showed that increasing
the fast-switching cycle time decreases the accuracy of empirical
phase coefficients in only a gradual manner.

Secondly, it can be seen that the empirical coefficients for Channels
3 and 4 are somewhat different to the `best-fitting' coefficients
obtained in Section 4 and Table 2 of \cite{ALMANikolic587}. The reason
is that best-fitting coefficients of \cite{ALMANikolic587} are found
by minimising the sums of squares of the residual phase errors while
in this paper I try to measure the true correlation between the
measured phase and the sky brightness fluctuation. Because there are
sources of error in both of these quantities, simply minimising the
squares of the residuals will lead to an underestimate of the
correction coefficients. There is a good discussion of this effect by
\cite{ALMAHoldaway515}.

\begin{figure*}

\begin{tabular}{cc}
  Without empirical coefficients & With empirical coefficients\\
  \includegraphics[clip,width=0.49\linewidth]{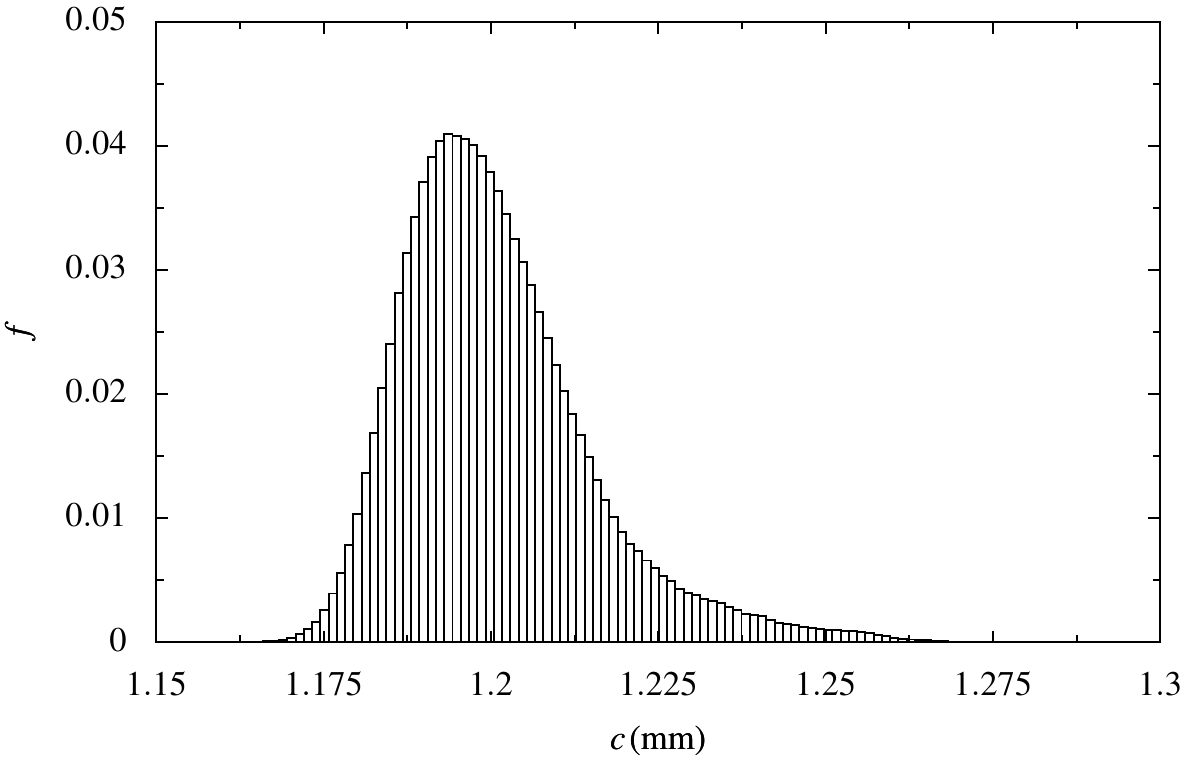}&
  \includegraphics[clip,width=0.49\linewidth]{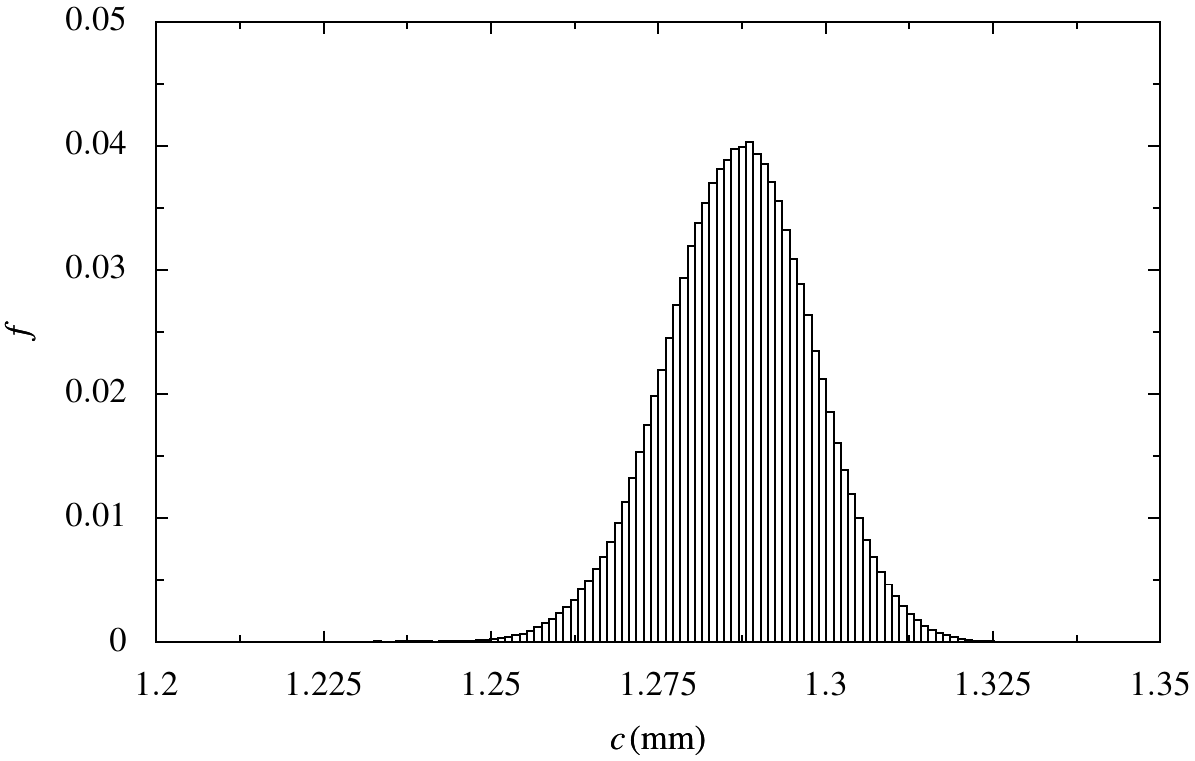}\\
  \includegraphics[clip,width=0.49\linewidth]{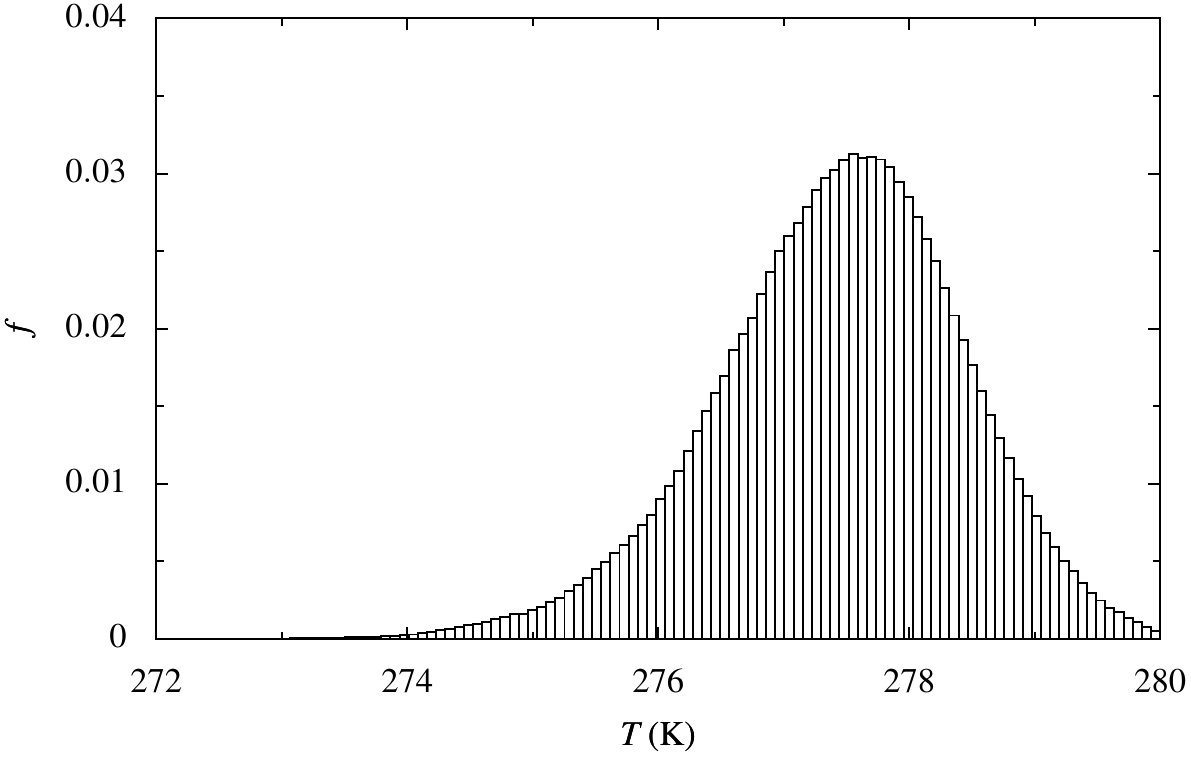}&
  \includegraphics[clip,width=0.49\linewidth]{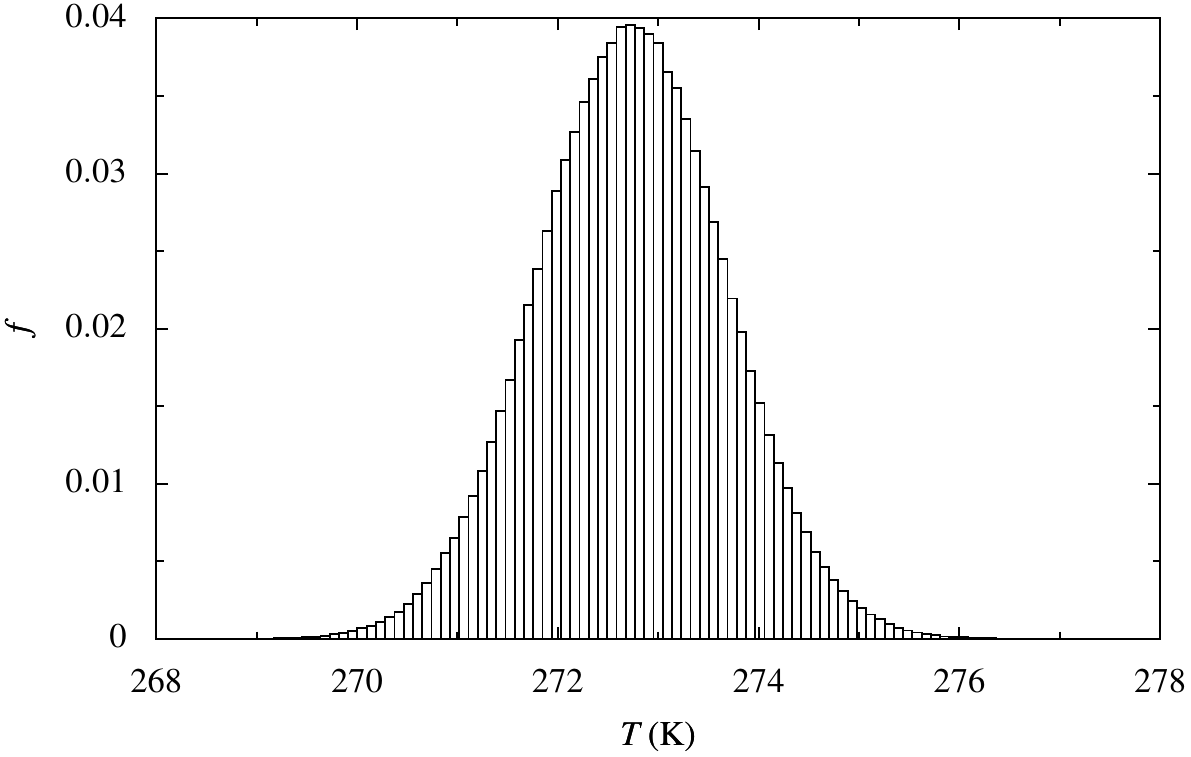}\\
  \includegraphics[clip,width=0.49\linewidth]{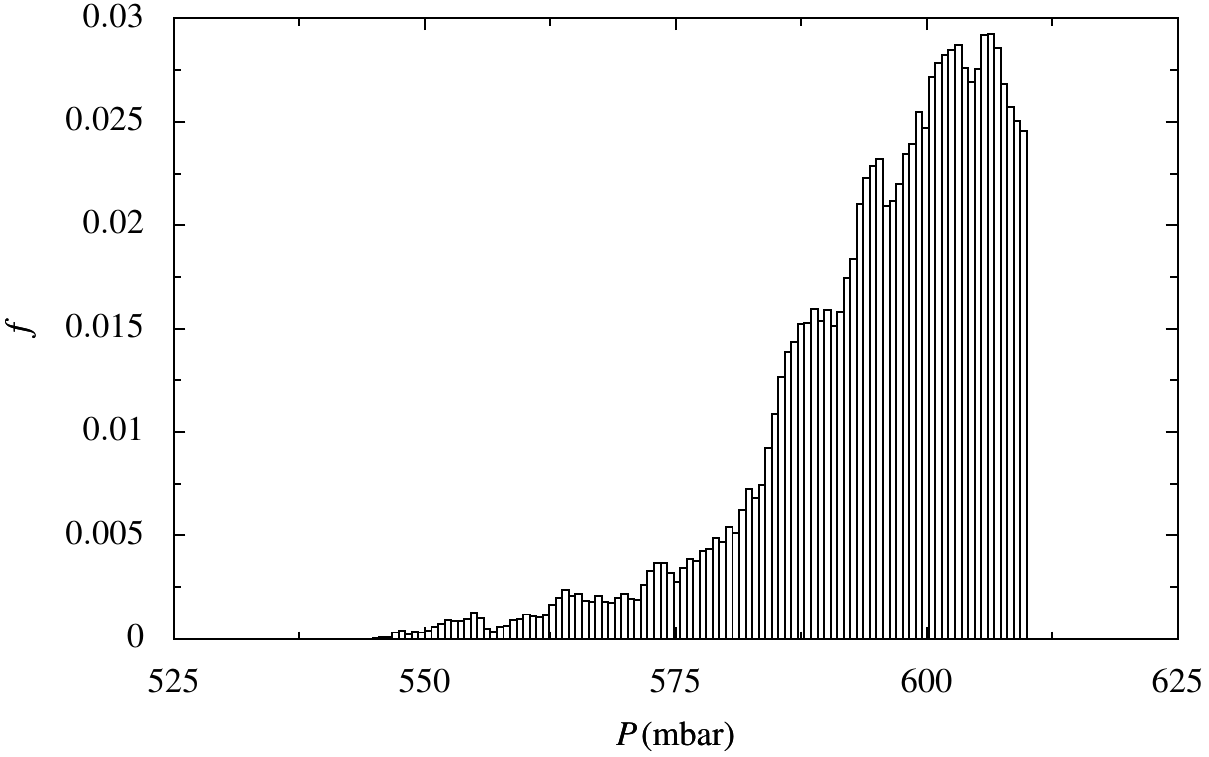}&
  \includegraphics[clip,width=0.49\linewidth]{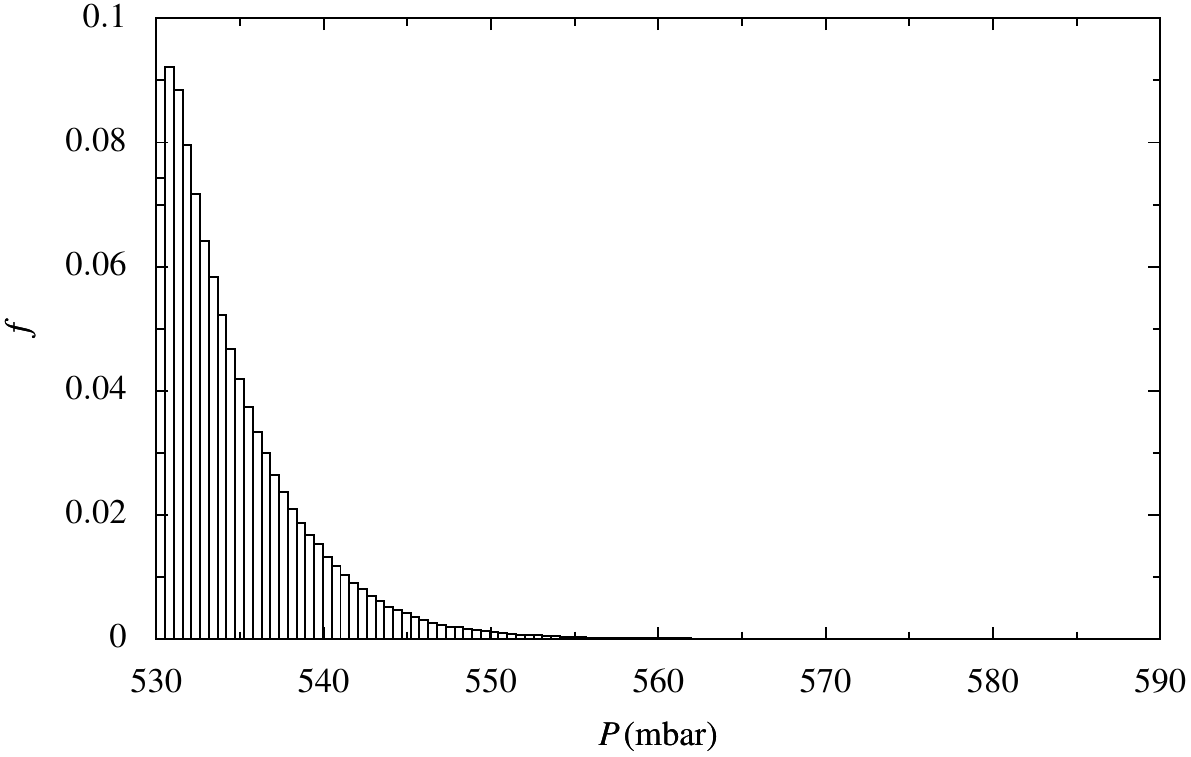}
\end{tabular}
\caption{Illustration of the retrieval of model parameters using as
  input only the absolute sky brightness (left column) and the
  absolute sky brightness and the empirical coefficients for channels
  3 and 4 (right column) for SMA observation on 17 February 2006. The
  absolute sky brightness was taken from the middle of the observation
  scan, while empirical coefficients were estimated for a few minutes
  around the middle of the scan.  In this case I assigned an error of
  4\% to the measurements of the empirical coefficients, and the error
  on the absolute sky measurement is as before assumed to be 1\,K.}
\label{fig:sma-emp-ret}
\end{figure*}

\begin{figure*}
\begin{tabular}{cc}
  Without empirical coefficients & With empirical coefficients\\
  \includegraphics[clip,width=0.45\linewidth]{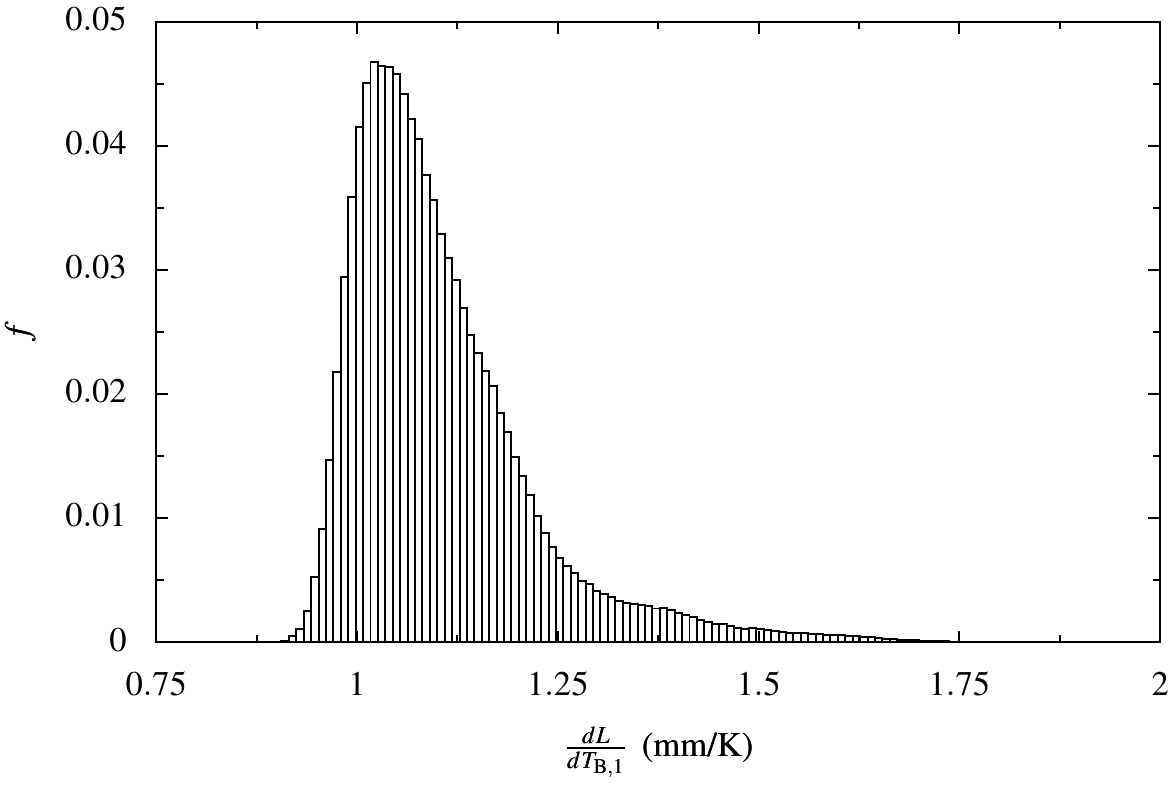}&
  \includegraphics[clip,width=0.45\linewidth]{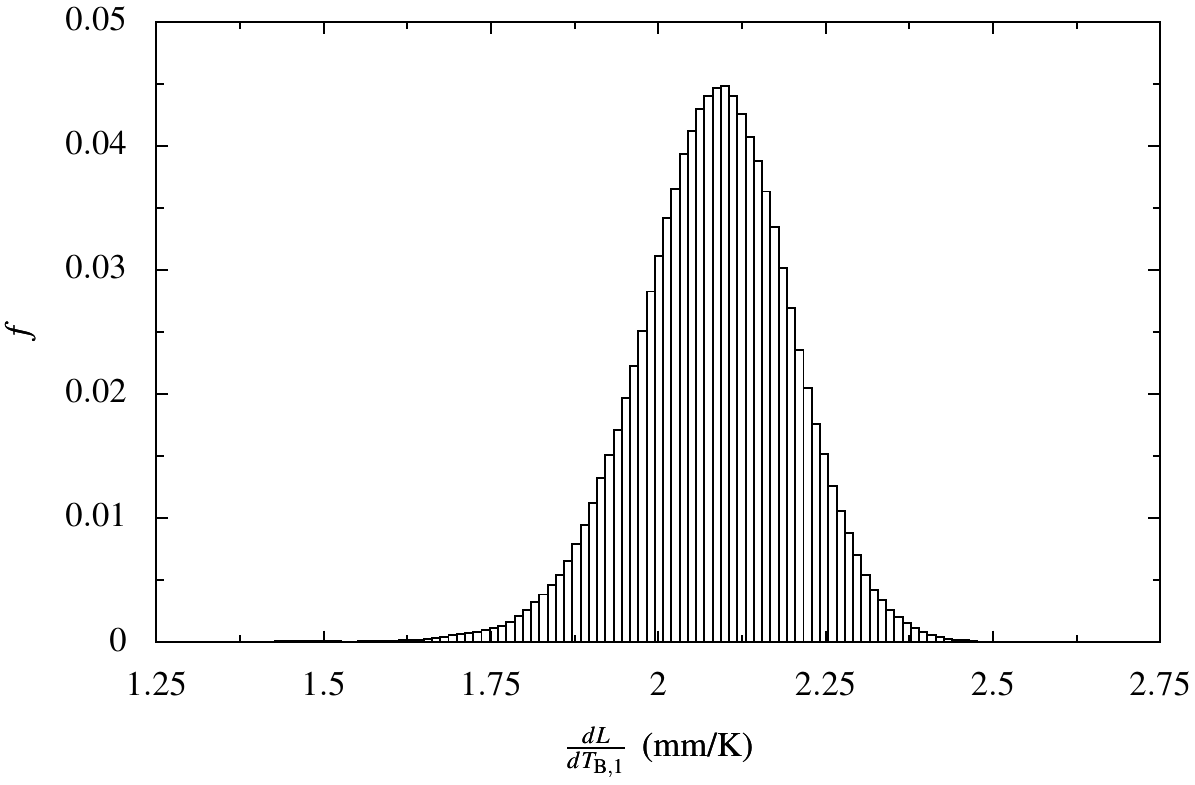}\\
  \includegraphics[clip,width=0.45\linewidth]{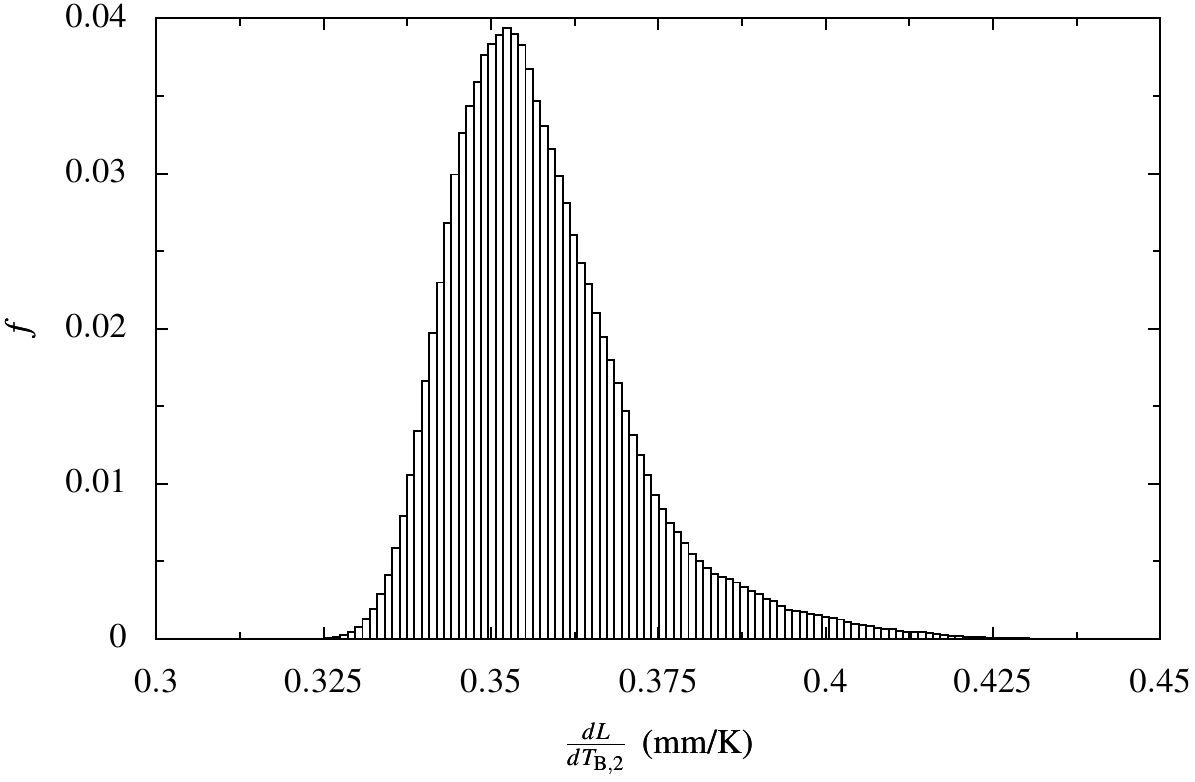}&
  \includegraphics[clip,width=0.45\linewidth]{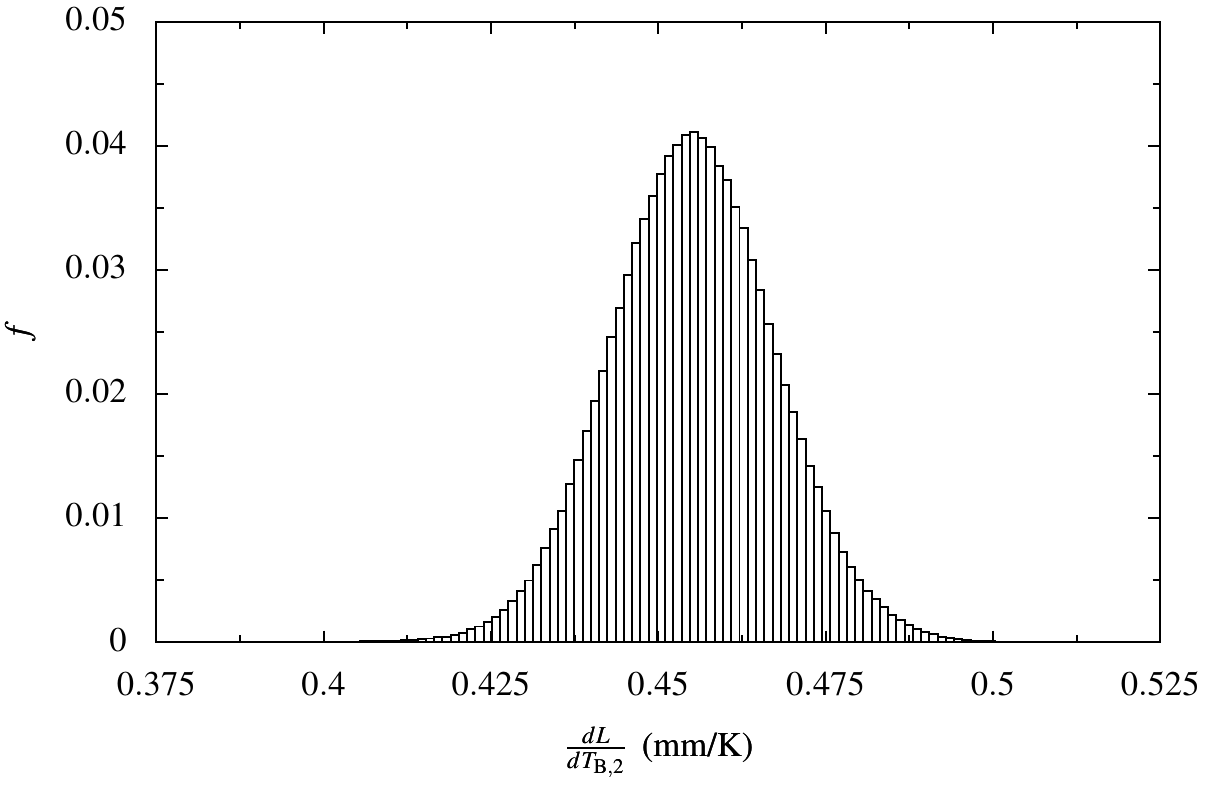}\\
  \includegraphics[clip,width=0.45\linewidth]{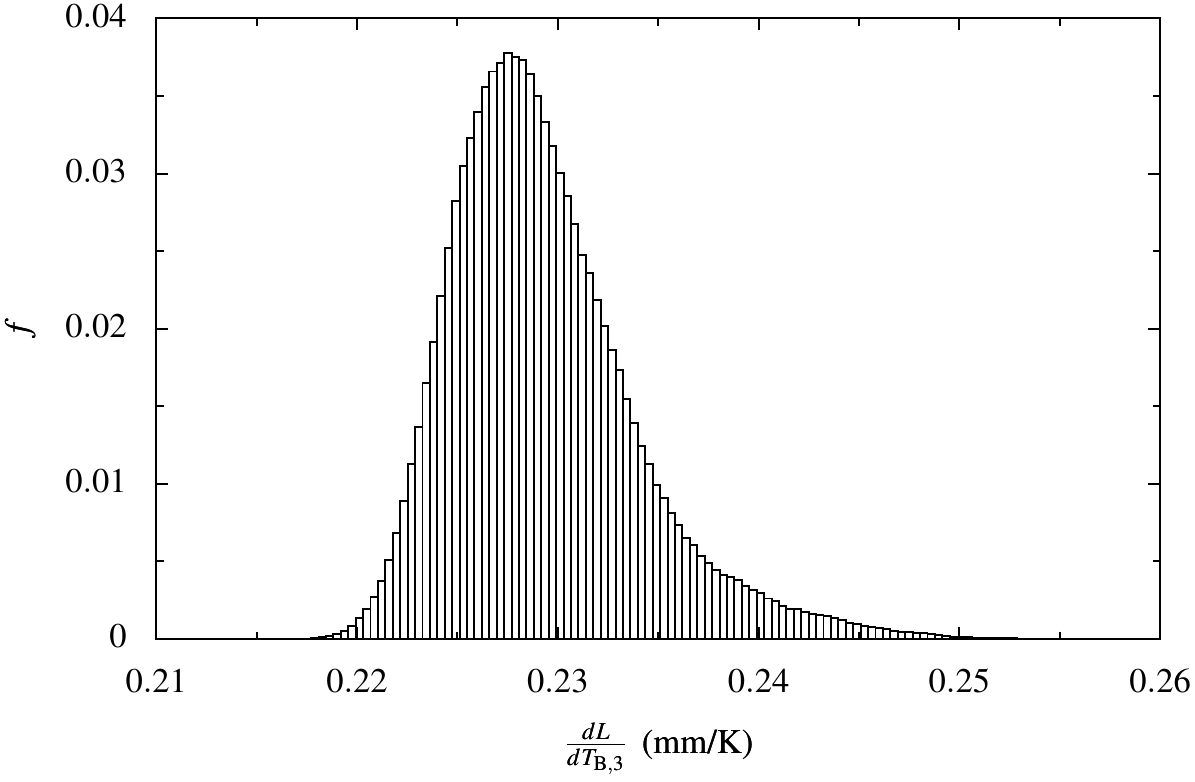}&
  \includegraphics[clip,width=0.45\linewidth]{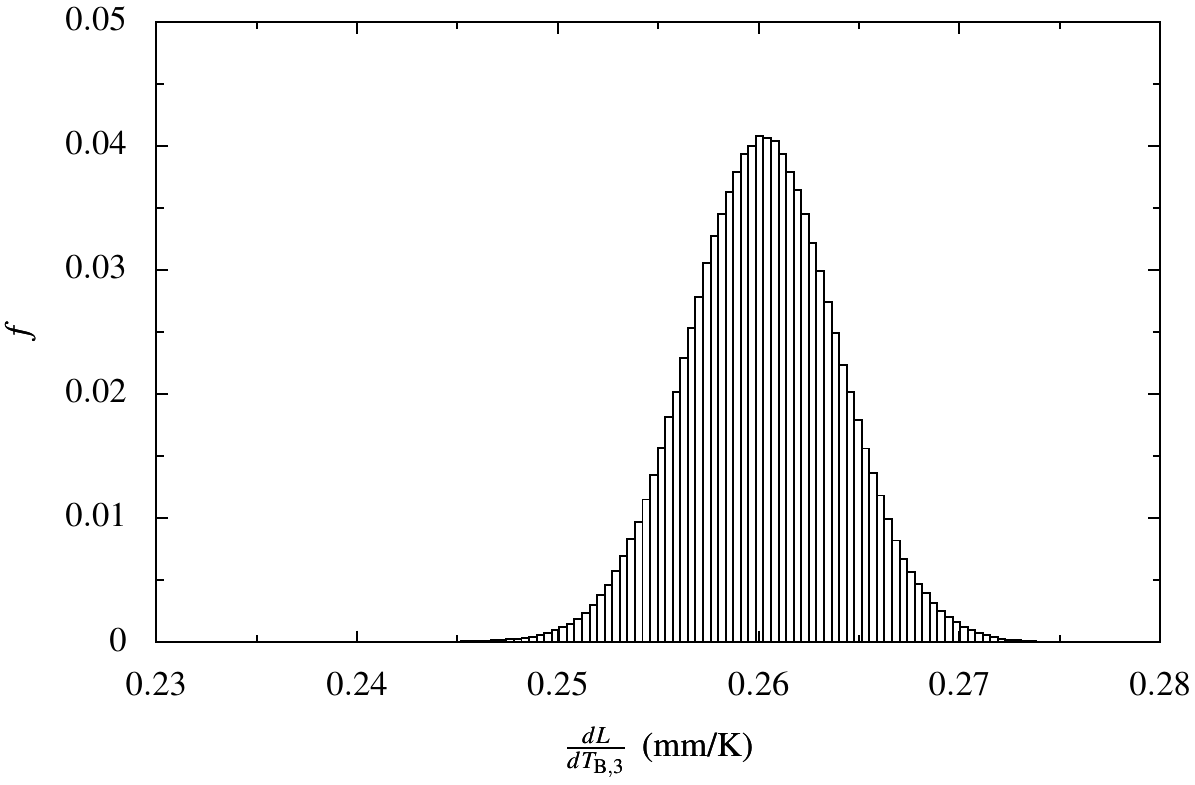}\\
  \includegraphics[clip,width=0.45\linewidth]{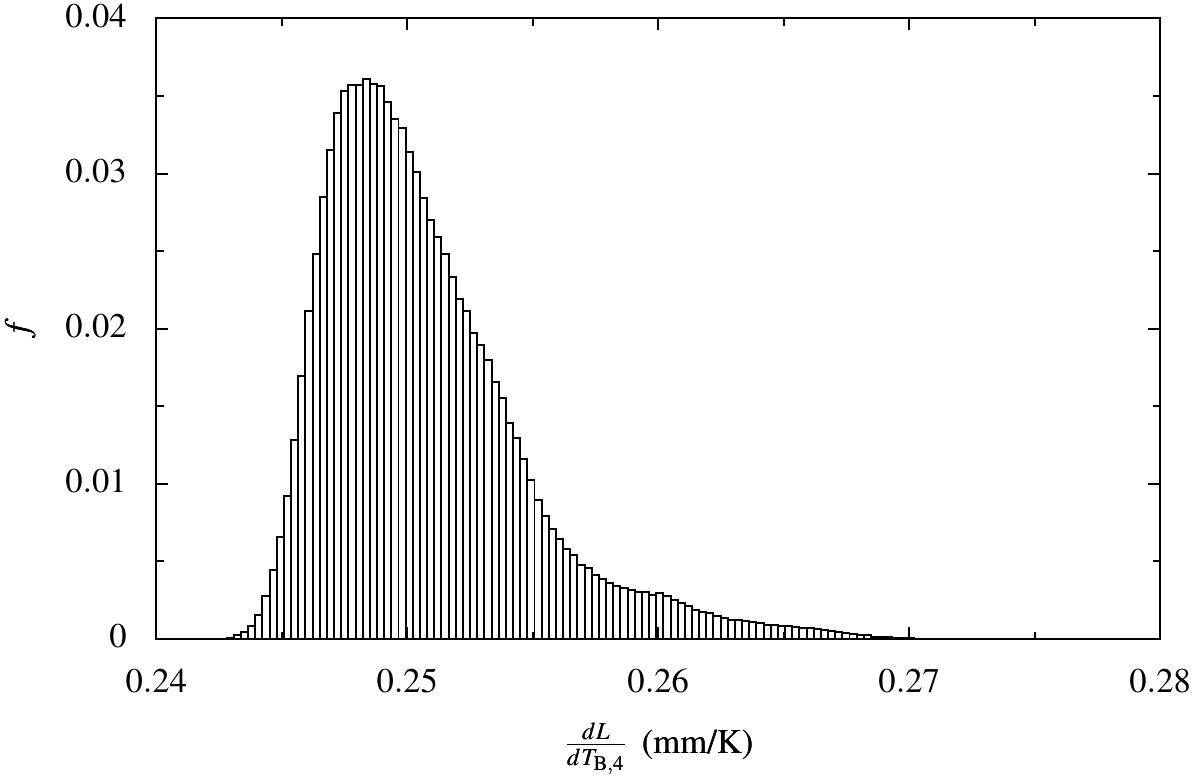}&
  \includegraphics[clip,width=0.45\linewidth]{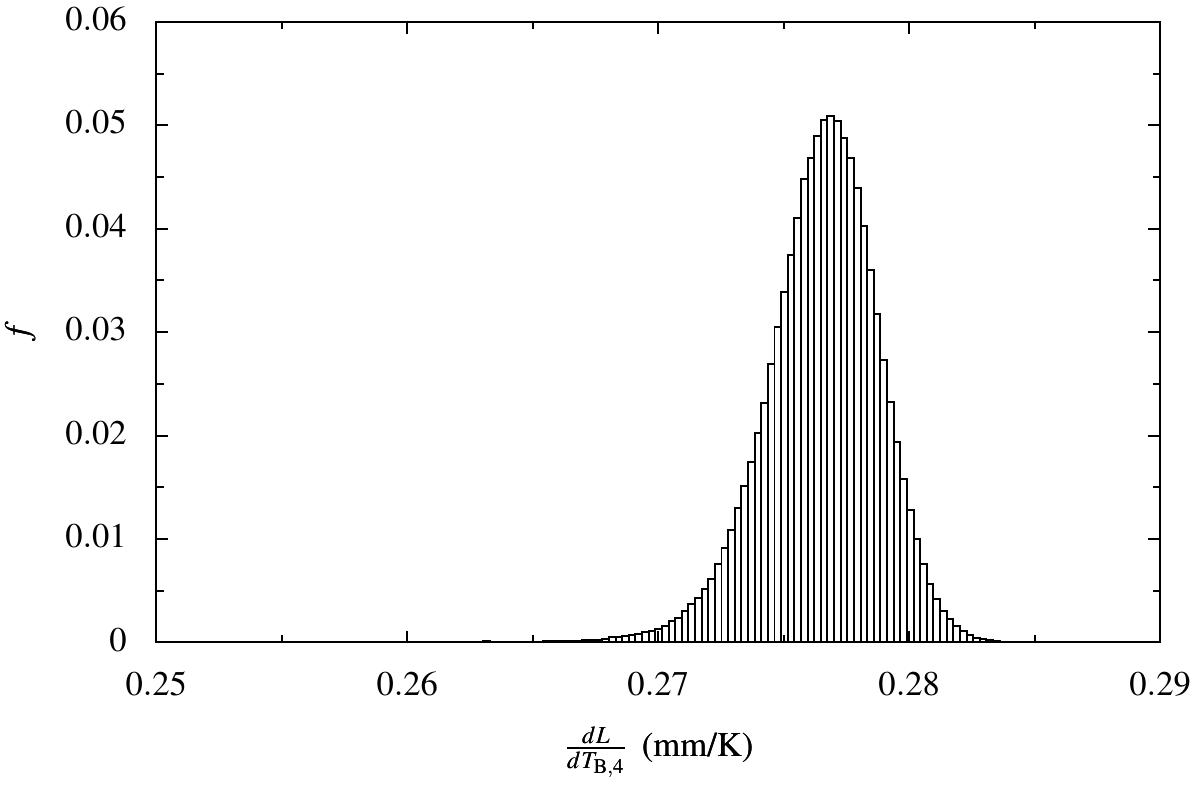}\\
\end{tabular}
\caption{Inferred phase correction coefficients corresponding to
  Figure~\ref{fig:sma-emp-ret}, i.e., based on only the absolute sky
  brightness (left column) and on the absolute sky brightness and the
  empirical coefficients for channels 3 and 4 (right column). }
\label{fig:sma-emp-dtdls}
\end{figure*}

The final part of sections shows the results of the joint analysis of
the absolute sky brightness and the empirical phase correction
coefficients (as discussed in \ref{sec:constr-model-param}) to place
constraints on the model parameters. For this analysis I took the
absolute sky brightness observed in the middle of the SMA observation
shown in Figure~\ref{fig:smadata}. The elevation of this time was 22
degrees and the source was setting rapidly. For this reason, I
computed the empirical coefficients for a shorter time around the
middle of the scan only. As described above, the results for channels
1 and 2 were not usable in these conditions, while the values of the
coefficients for channels 3 and 4 were 313 and 324\,\micron/K
respectively.

Because in this memo I again do not take into consideration the
dispersive effects of the water vapour, these figures need to be
adjusted for the expected dispersive phase at the frequency of
observation (240\,GHz in this case). As in \cite{ALMANikolic587} I
take this correction factor to be 1.05, resulting in coefficients of
298 and 309\,\micron/K for Channels 3 and 4 respectively.  These
empirical coefficients were then analysed together with the four
absolute brightness which in this case were 268.3, 247.4, 200.4,
135.9\,K for channels 1 to 4 respectively. The errors assigned on the
absolute sky brightness were again 1\,K while I assigned error of 4\%
on the empirical coefficients.

The marginalised distributions of the model parameters produced by
this analysis are shown in Figure~\ref{fig:sma-emp-ret} side-by-side
with the results of an analysis based on the absolute sky brightness
only (essentially the same as shown in the previous memo). From this
comparison, it can be seen that significantly different distributions
are inferred for the all three of the model parameters. In particular,
even the inferred water vapour column is moved to a significantly
higher value, from about 1.2 to 1.3\,mm of water vapour at
zenith. Also the pressure distribution is moved from being mostly
constrained by the high-end of the prior range to being mostly
constrained by the low-end of the prior range.

These changes in model parameters when the observations of the
empirical coefficients are introduced into the analysis should of
course be interpreted as the necessary moves to reconcile the
predicted coefficients with those actually observed. The fact that the
two model parameter distributions do not overlap does, however,
suggest that the inclusion of the observed empirical coefficients is
also correcting for a modelling and/or calibration error and not just
for degeneracies present in the problem as can be seen in
Figure~\ref{fig:modelconstraint}.

Finally, the marginalised distributions of the predicted phase
correction coefficients are shown in Figure~\ref{fig:sma-emp-dtdls},
again in a side-by-side format with the retrieval based just on
absolute sky brightnesses on the left and the retrieval based on the
empirical coefficient observations \emph{and\/} the absolute sky
brightness on the right. As expected, the distributions of the phase
correction coefficients are narrower (and closer to Gaussian) when the
observed coefficients are included the analysis, both for the channels
with observational constraints (channels 3 and 4) and for the other
channels (1 and 2). The second point that should be noted is that the
marginalised distribution of channels 3 and 4 are not centred on the
observed values that were entered into the analysis. The reason is
that the assigned 4\% error allows for variation in the retrieves
values while the observed absolute sky brightness temperatures clearly
prefer a lower value of the coefficients.

\section{Summary and Discussion}

The results presented above may be divided into two main
parts. Firstly, in Section~\ref{sec:obssacc}, the limits due to
thermal noise on observational determination of the phase correction
coefficients are calculated. Because only thermal noise in the
receivers is considered, this analysis should be regarded a best-case
scenario; in practice effects such as electronic and/or mechanical
phase drifts will always be become important on a long enough
timescale. These effects are by their nature very much less
predictable than thermal noise in receivers and realistic estimates of
their magnitude will be possible only once the commissioning of ALMA
at the array operations site begins.

In terms of the technique of estimating the empirical phase correction
coefficient from observed data two points from this section should be
noted:
\begin{enumerate}
  \item Best-fitting linear relationship should be computed taking
    into account errors on both of the coordinates in the fit (i.e.,
    both on the measured radiometric fluctuation and measured phase).
  \item When combining estimates from baselines in a two-dimensional,
    array, they must be correctly weighted by their error estimates 
\end{enumerate}
The last of these is important because errors on estimates on short
baselines are \emph{much\/} greater than the estimates from long
baselines.

The main implications of results in Section~\ref{sec:obssacc} is that
it should be possible to accurately and relatively quickly estimate
the empirical phase correction coefficients. Even with a single
baseline array, it should be possible to get to about 1\% accuracy in
about 200\,seconds.  With full ALMA in a medium configurations
accuracies of order of 0.1\% could be reachable if non-thermal effects
do not become important.

These predicted accuracies should be compared to results of
\emph{a-priori\/} modelling of the phase correction coefficients as
for example given in the previous memo \citep{ALMANikolic587}. There,
I estimated the error on the predicted phase correction coefficients
due to the uncertainties in the input parameters only, i.e., intrinsic
error on the measurement of absolute sky brightness with the WVRs and
the poorly known pressure/temperature of the water vapour in the
atmosphere.  The error estimates in \cite{ALMANikolic587} do not
therefore include modelling error. I found that if we do not have a
good constraint on temperature and or height of the water vapour, the
range of calculated coefficients spans about 8\%, while if there are
good constraints, the range spans around 3-4\%.

Therefore, according to the above calculations, relatively short
empirical measurements are likely to have smaller errors than the
\emph{a-priori} predictions. For this reason empirical observations
should be very useful for phase correction, at least until we learn
how to estimate the atmospheric parameters which are not well
constrained by absolute measurements from the WVRs (the parameter that
\emph{is\/} well constrained by the WVRs is of course the total water
vapour column). 

The second part of the memo (Sections \ref{sec:results} and
\ref{sec:smaillustration}) concerns the way empirically determined
phase correction coefficients are used in practice. 

The thesis put forward in this part of the memo is that the best way
of using these coefficients is to regard them as additional
observational measurement which are used to constrain a physical model
of the atmosphere, such as the simplified model presented in the first
memo of this series \citep{ALMANikolic587}. The advantages of this
approach are that:
\begin{enumerate}
  \item It is possible to easily and accurately transfer measured
    empirical coefficients to different conditions, such as different
    elevation or a part of the sky with slightly different pwv column
  \item Constraints in the underlying physical model can
    \emph{improve\/} the accuracy of the empirical measurements 
  \item The same constraints can also flag implausible measurements of
    the empirical coefficients
  \item It is possible to do \emph{quantitative\/} model selection,
    using all of the available information
\end{enumerate}
The existing algorithms presented in the previous memo can naturally
be extended for such an analysis and this has been already implemented
in the {\tt libAIR} library. 

Finally here are some caveats and directions for future work.  One of
the topics that has not been addressed in this memo are the effects of
dispersion. If they can be modelled accurately, there is no reason why
dispersive effects can not be taken into account in the calculation of
the predicted ${\rm d}L/{\rm d} T_{{\rm B}}$ and the subsequent
analysis can proceed as before.

A second topic which may be important is that of the so-called `dry'
fluctuations, that is, fluctuations in phase due to changes in the
density of dry air rather than quantity of water vapour. If these are
completely independent of the `wet' fluctuations then it will
certainly not be possible to correct for them, and they will manifest
themselves as an extra source of error when calculating the empirical
correction coefficients. There errors assigned to the path-fluctuation
coordinate used in the line-fitting procedure should take into account
this extra source of error.

It is likely however that the `dry' fluctuations are partially
correlated with the `wet' fluctuations, as for example predicted by
the Large Eddy Model simulations of \cite{ALMAStirling517}. In this
case the observed phase correction coefficients will be scaled by an
unknown factor from their predicted values. If this effect is
significant, it will likely be counter productive to let the inference
procedure try to match the observed coefficients. Rather, the scaling
should be introduced into the model as a further parameter and
Bayesian evidence calculation then used to determine the significance
of the dry fluctuations.

\section*{Acknowledgements}

I would like to thank Robert Laing for a careful reading of the
manuscript and suggestions for improving it and making it clearer.

This work is a part of the ALMA Enhancement programme at the
University of Cambridge funded by the European Union Framework
Programme 6. More information on the programme is available at
\url{http://www.mrao.cam.ac.uk/projects/alma/fp6/}.

I would like to thank the team involved in construction and the
testing at the SMA of the ALMA prototype water vapour radiometers:
P.~G. Anathasubramanian, R.~E. Hills, K.~G.Isaak, M. Owen,
J.~S. Richer, H. Smith, A.~J. Stirling, R.~Williamson, V.~Belitsky,
R. Booth, M. Hagstr\"om, L. Helldner, M. Pantaleev, L.~E.~Pettersson,
T.~R. Hunter, S. Paine, A. Peck, M.~A. Reid, A.~Schinckel and
K.~Young.

\bibliographystyle{mn2e} 
\bibliography{wvrretrieval.bib}

\vskip 1cm

{\flushleft \footnotesize Internal revision control information: \bzrrevno (\bzrdate)}

\label{lastpage}
\end{document}